\UseRawInputEncoding
\documentclass[usenatbib,useAMS,usedcolumn]{mnras}

\usepackage{epsfig}
\usepackage{times,pdflscape}
\pdfoutput=1
\hypersetup{pdfauthor={H.L. Liu},
            bookmarksnumbered=true}

\newcommand{\cm}{\ensuremath{\mbox{~cm}}}

\newcommand{\pcmcu}{\ensuremath{\cm^{-3}}}

\newcommand{\Hii}{H\textsc{ii}}

\newcommand{\vel}{km\,s$^{-1}$}
\newcommand{\msun}{$M_{\odot}$}
\newcommand{\lsun}{$L_{\odot}$}
\newcommand{\um}{$\mu$m}
\newcommand{\cmcm}{cm$^{-2}$}
\newcommand{\egcite}{\citep[e.g.,][]}

\newcommand{\chiiioh}{CH$_3$OH}
\newcommand{\hciiin}{HC$_3$N}

\newcommand{\halpha}{H$40_{\alpha}$}
\newcommand{\chthocho}{CH$_3$OCHO}
\newcommand{\chthcho}{CH$_3$CHO}
\newcommand{\chii}{H/UC-H\textsc{ii}}
\newcommand{\uchii}{UC-H\textsc{ii}}
\newcommand{\hchii}{HC-H\textsc{ii}}
\newcommand{\CHMC}{s-cHMC}
\newcommand{\PCHMC}{w-cHMC}

\graphicspath{{./0figs/}}

\begin{document}

\title[ ATOMS catalogues of high-mass star-forming cores
] 
{ 
ATOMS: ALMA Three-millimeter Observations of Massive Star-forming regions –- III. Catalogues of candidate hot molecular cores and Hyper/Ultra compact H{\sc ii} regions
}


\author[H.-L. Liu et al.]{
Hong-Li Liu,$^{\star 1,4,2}$
Tie Liu,$^{\star 2,3}$
Neal J. Evans,$^{5,6}$
Ke Wang,$^{7}$
Guido Garay,$^{8}$
Sheng-Li Qin,$^{1}$
\newauthor
Shanghuo Li,$^{6}$
Amelia Stutz,$^{4,9}$
Paul F. Goldsmith,$^{10}$
Sheng-Yuan Liu,$^{11}$
Anandmayee Tej,$^{12}$
\newauthor
Qizhou Zhang,$^{13}$
Mika Juvela,$^{14}$
Di Li,$^{15,16,17}$
Jun-Zhi Wang,$^{2,18}$
Leonardo Bronfman,$^{8}$
\newauthor
Zhiyuan Ren,$^{15}$
Yue-Fang Wu,$^{19}$
Kee-Tae Kim,$^{6,20}$
Chang-Won Lee,$^{6,20}$
Ken'ichi Tatematsu,$^{21}$
\newauthor
Maria. R. Cunningham,$^{22}$
Xun-Chuan Liu,$^{19}$
Jing-Wen Wu,$^{15}$
Tomoya Hirota,$^{21}$
\newauthor
Jeong-Eun Lee,$^{23}$
Pak-Shing Li,$^{24}$
Sung-Ju Kang,$^{6}$
Diego Mardones,$^{8}$
Isabelle Ristorcelli,$^{25}$
\newauthor
Yong Zhang,$^{26}$
Qiu-Yi Luo,$^{2}$
L. Viktor Toth,$^{27}$
Hee-weon Yi,$^{23}$
Hyeong-Sik Yun,$^{23}$
\newauthor
Ya-Ping Peng,$^{28}$
Juan Li,$^{2,18}$
Feng-Yao Zhu,$^{2}$
Zhi-Qiang Shen,$^{2,18}$
Tapas Baug,$^{7,29}$
\newauthor
Lokesh Dewangan,$^{30}$
Eswaraiah Chakali,$^{15}$
Rong Liu,$^{15}$
Feng-Wei Xu,$^{19}$
Yu Wang,$^{19}$
\newauthor
Chao Zhang,$^{31,15}$
Jinzeng Li,$^{15}$
Chao Zhang,$^{1}$
Jianwen Zhou,$^{15}$
Mengyao Tang,$^{1}$
Qiaowei Xue,$^{1}$
\newauthor
Namitha Issac,$^{12}$
Archana Soam,$^{32}$
Rodrigo H. \'Alvarez-Guti\'errez$^{4}$
\\
Affiliations are listed at the end of the paper}

\date{Accepted 7 May 2021; Received 7 May; in original 22 December 2020}

\pagerange{\pageref{firstpage}--\pageref{lastpage}} \pubyear{2021}

\maketitle

\label{firstpage}

\begin{abstract} 

We have identified 453 compact dense cores in  3\,mm continuum emission maps in the ATOMS (ALMA Three-millimeter Observations of Massive Star-forming regions) survey, and compiled three catalogues of high-mass star forming cores.
One catalogue,  referred to as \chii\ catalogue, includes 89 cores that enshroud hyper/ultra compact (H/UC) H{\sc ii} regions as characterized by associated compact \halpha\ emission. 
A second catalogue, referred to as pure s-cHMC, includes 32 candidate Hot Molecular Cores (HMCs) showing rich spectra ($N\ge20$ lines) of complex organic molecules (COMs) but not associated with \chii\ regions.
The third catalogue, referred to as pure w-cHMC, includes 58 candidate HMCs with relatively low levels of COM richness and not associated with \chii\ regions. These three catalogues of dense cores provide an important foundation for future studies of the early stages of high-mass star formation across the Milky Way. We also find that nearly half of \chii\ cores are candidate HMCs. From the number counts of COM-containing and \chii\ cores, we suggest that the duration of high-mass protostellar cores  showing chemically rich features is at least comparable to the lifetime of \chii\ regions. For cores in the \chii\ catalogue, the width of the \halpha\ line increases as the core size decreases, suggesting that the non-thermal  dynamical and/or pressure line-broadening mechanisms dominate on the smaller scales of the \chii\ cores.

\end{abstract} 


\footnotetext[1]{E-mail: hongliliu2012@gmail.com, and liutie@shao.ac.cn}

\section{Introduction}
\label{sec:intro}

High-mass stars ($M_{\star}>8$\,\msun) play a crucial role in many astrophysical processes, for example from the formation of solid material
in the early Universe \citep{Dun03} to the influence on the evolution of their host galaxies and future generations of star formation in their natal 
molecular clouds \egcite{Ken05,Urq14}. Despite its importance on different (time and spatial) scales, 
high-mass star formation remains much less understood than low-mass star formation due to the observational challenges presented by increased distances, rarity, opaque surroundings, and short formation timescales \egcite{Zin07,Tan14,Mot18}. In addition, high-mass protostars are embedded in a rich cluster environment \egcite{de 05}, which makes it very difficult to disentangle the physical formation processes of high-mass stars and numerous low-mass cluster members 

Despite the observational difficulties mentioned above, it has been known for the past 15 years that
high-mass stars spend a considerable fraction of their lifetime ($>10\%$) embedded in their natal molecular clouds \egcite{van04}. 
From an observational perspective, this embedded phase can be generally subdivided into four different groups of
objects that represent different evolutionary stages. \textbf{1)} 
Massive dense cores (MDCs, M $\ga30$\,\msun\ within $\sim 0.1$\,pc),   They are generally the nurseries of high-mass star formation \egcite{Zha09,Wan11,Zha15,Mot18,San19,Li 19,Svo19}, and 
perhaps analogous to the prestellar cores in low-mass star formation regions if they are completely starless. \textbf{2)} High-mass protostellar objects (HMPOs). They are very luminous, with bolometric luminosities above $10^3$\,\lsun, and deeply embedded within massive envelopes having centrally-peaked temperature and density distributions. At this stage, a (pseudo) disk is formed to transfer onto the HMPOs the infalling material from the envelope \egcite{Beu02,McK03,Ces07}. \textbf{3)} Hot molecular cores (HMCs). They correspond to the compact ($<0.1$\,pc), dense ($\ge10^5-10^8$\,\pcmcu),
massive ($\sim 100$\,\msun), and hot ($\ga100$\,K) molecular material that is 
 radiatively heated intensively \egcite{Kur00,Rat11, Hos09}. Consequently, the intense radiative heating enriches the gas phase with many complex organic molecules (COMs) such as ${\rm CH_3OCH_3}$ and ${\rm CH_3OCHO}$.
\textbf{4)} Hyper/ultra compact (H/UC)-\Hii\ regions. The \uchii\ regions refer to those with sizes $\le0.1$\,pc, densities $>10^4$\,\pcmcu, and emission 
measures $\ge10^7$\,pc~cm$^{-6}$ \egcite{Woo89,Hoa07}.
The \hchii\ regions, which are a newly identified class of compact \Hii\ regions after the discovery of \uchii, correspond to those with sizes  $\le0.05$\,pc,  densities $>10^6$\,\pcmcu,  and  
emission  measures $\ge10^7$\,pc~cm$^{-6}$ \citep{Kur05, Hoa07,Yan19,Yan20}.

During the past decades, studies of the embedded phase of high-mass star formation have made much progress. After a large number of IRDCs were revealed 
by Infrared Space Observatory (ISO, \citealt{Per96}), Midcourse Space Experiment(MSX, \citealt{Ega98}), and {\it Spitzer} \citep{Per09}, 
several surveys in (sub)millimeter continuum and spectral lines followed 
\egcite{Rat06, Rag06,  Sim06, Rat10, Dir15, San19}, 
and have established that massive IRDCs harbor the early, embedded stages of high-mass star and cluster formation. 
In particular, the recent ALMA 1\,mm survey ASHES led by \citet{San19}, which was designed to mosaic 12 massive ($>500$\,\msun), cold ($\le15$\,K), 3.6-70\,\um\ dark-cloud clump 
at the resolution $\sim 4000$\,AU and mass sensitivity better than $0.2$\,\msun,
revealed a large population of low-mass($<1$\,\msun)  cores but no  high-mass ($>30$\,\msun) prestellar cores on the 0.01--0.1\,pc scales.
With detailed analysis, the authors also put several strong observational constraints on cloud fragmentation, the clustering mode of dense cores, and the dynamical collapse scenario at the early stages of high-mass star formation.

As for the HMPO and HMC stages, both have many observational similarities except for the abundance of COMs; for example, they both
are accompanied by massive molecular outflows, infall and even rotation\egcite{Zha01,Zha05,Beu02,van04,Liu17}. Some important physical processes including mass accretion and disk rotation related to the HMPOs can therefore be 
understood from the investigation into the HMCs since the latter is generally accompanied by rich molecular line emission
that bears both physical and chemical imprints of high-mass star formation\egcite{Ket10,Qiu12}.
Focusing on the HMCs, we find that most of the efforts so far have been toward individual sources 
or small samples through single-dish and/or interferometric observations in the 
(sub)millimeter regime \egcite{van04,Beu07,Qin08,Pur09,Ces10,Fur11,Rat11,Xu 13,Her14,Qin15,Sil17,Cse19,Col20,Bel20,Sat20}.
Although these studies have demonstrated the HMC as an ideal laboratory in which to search for infall, outflow, and rotation motions related
to high-mass star formation,  these important physical
processes are still far from being well understood mainly due to the lack of deep, interferometric (sub)millimetre observations toward
a sufficiently large sample of HMCs. For the \chii\ stage, the situation improves; 
there are several surveys that serve to characterize the \chii\ regions, such as IRAS warm dust continuum 
at far-IR wavelengths \citep{Woo89,Kur94}, CORNISH 5\,GHz radio continuum \citep{Hoa12,Pur13}, 
and ATLASGAL 870\,\um\ cold dust continuum \citep{Sch09,Urq13} surveys. 
These surveys help define observationally the \chii\ regions reasonably well in terms of their sizes, densities, and emission measures (see above).

In addition, some \chii\ regions are likely still accreting mass, despite the high pressure of ionized gas\egcite{Sol05,Ket07,Ket08}. However,  the role that  \chii\ regions play in the early stages of high-mass star formation
remains to be understood. Hopefully, the \chii\ regions can light up their immediate surroundings and allow the investigation of the properties (e.g., 
density distribution, and velocity field) of the surrounding region in which high-mass stars are formed \egcite{Hoa07}. The immediate vicinities of \chii\ regions
have been the subject of the (sub)millimeter observation studies, however, the lack of high-angular resolution observations has impeded anything like complete
understanding of \chii\ region physics.

In this paper, the third in a series from ATOMS\footnote{ATOMS: ALMA Three-millimeter Observations of Massive Star-forming regions survey} 
(e.g., \citealt{Liu20}, \citealt{Liu20b}, hereafter, Paper\,I and Paper\,II, respectively), 
we take advantage of deep, high-resolution ALMA 3\,mm observations (see below) of the
ATOMS survey. 
In Paper\,I, we introduced the main scientific goals of the survey through a case
study for the G9.62+0.19 complex, including to systematically investigate the spatial distribution of various dense gas tracers in a large sample of Galactic massive star-forming clumps, to study the roles of stellar feedback in star formation, and to characterize filamentary structures inside massive clumps. In paper\,II, we studied the star formation scaling relations inferred from
different dense gas tracers, and suggested that both the main and isotopologue lines (i.e., HCO$^+$/H$^{13}$CO$^+$, HCN/H$^{13}$CN ) are good tracers of the total mass of dense gas in Galactic molecular clumps, and that the large optical depths of the main lines do not affect the interpretation of the slopes in star formation relations.

With the ATOMS survey data, in this paper we aim to
establish catalogues of a large sample of both candidate HMCs and \chii\ regions as
a crucial foundation for future studies for the early stages of high-mass star formation.
The ATOMS survey targeted a large sample of 146 IRAS clumps \citep{Bro96,Liu16}, 
which have masses 5.6 to $2.5\times10^5$\,\msun\ with a median value of $1.4\times10^3$\,\msun, radii 0.06 to 4.26\,pc with a median value 
of $0.86$\,pc, and bolometric luminosities 16 to $8.1\,\times10^6$\,\lsun\ with a median value of $5.7\,\times10^4$\,\lsun\ (see Table\,A1 of  Paper\,I).
All the targets except for I08076–3556 and I11590–6452 have bolometric luminosities greater than $10^3$\,\lsun, and actually more than 
90\% of them are high-mass star-forming regions (see Paper\,I, and Paper\,II). 
In addition, the ATOMS targets are located in the first and fourth Galactic Quadrants of the inner Galactic plane ($-80$\degr$<l<40$\degr, $|b|<2$\degr) at distances
0.4 to 13.0\,kpc (the corresponding Galactocentric distances 0.5 to 12.7\,kpc, see Appendix\,\ref{sec:app-kin-dist} for the distance calculations).  
Twenty-seven distant ($d \ge 7$\,kpc) sources are either close to the Galactic Centre or mini-starbursts (like W49N/I19078+0901), representing extreme environments for star formation.
Overall, the ATOMS survey contains a diversity of objects suitable for studying the early stages of high-mass star formation, especially HMC and \chii\ stages, with different physical conditions -- including densities and  luminosities -- and in different environments, across a large range of Galactocentric distances.

This paper is organized as follows: Section\,2 gives a brief description about the ALMA observations of the ATOMS survey, Section\,3 presents 
the results of the extraction and search of the candidate HMC and \chii\ cores, Section\,4 presents a  discussion, and Section\,5 includes  
the conclusions and a summary of our results.

\section{ALMA Observations}
\label{sec:observations}
We make use of the ATOMS survey data (Project ID: 2019.1.00685.S; PI: Tie Liu, see Paper\,I and Paper\,II). The observations 
were conducted towards the 146 IRAS clumps in the single-pointing mode with both the Atacama Compact 7-m Array (ACA; Morita Array) 
and the 12-m array (C43-2 or C43-3 configurations) in band\,3. Eight spectral windows (SPWs) were tuned to cover 11 commonly-used lines including the dense gas tracers (e.g., HCO$^+$, 
HCN, and their isotopes), HMC tracers (e.g., CH$_3$OH, and HC$_3$N), shock tracers (e.g., SiO, and SO), and ionized gas tracers (H$_{40}\alpha$). The basic parameters (e.g., rest frequency, transition,
critical density, and upper energy temperature) of these lines can be found in Table\,2 of Paper\,I. The SPWs\,1--6 \footnote{The frequency ranges of the eight spectral windows (SPWs) are [86.311, 86.369]\,GHz for SPW\,1, [86.725, 86.784]\,GHz for SPW\,2, 
[86.818, 86.876]\,GHz for SPW\,3, [87.288, 87.346]\,GHz for SPW\,4, 
[88.603, 88.661]\,GHz for SPW\,5, [89.159, 89.218]\,GHz for SPW\,6,
[97.530, 99.404]\,GHz for SPW\,7, [99.468, 101.341]\,GHz for SPW\,8.} at the lower side band, with spectral resolutions
of $\sim 0.2-0.4$\,\vel\ were chosen to resolve the line profiles in order to investigate the kinematics within high-mass star forming clumps, while the SPWs\,7--8 at the upper side band, 
each with a broad bandwidth of 1875\,MHz at a spectral resolution of $\sim 1.6$\,\vel, were chosen for  continuum emission and line scan observations. Note that
the original spectral resolution of the SPWs\,7--8 is $\sim 1.6$\,\vel instead of  $\sim 3$\,\vel\ as given in Paper\,I.

The ACA and 12\,m-array data were calibrated and imaged separately with the CASA software package version 5.6 \citep{McM07}, where 
the continuum image centred at $\sim93.8$\,GHz was cleaned in an aggregated $\sim 4$\,GHz frequency bandwidth free of the line-emission.  More details on the data reduction can be found in Paper\,I and Paper\,II.
 As we focus in this work on the very dense, compact HMCs and \chii\ regions, in what follows we will only consider the 12\,m-array data and the analysis is focused on the primary-beam corrected data. The reduced 12m\,-array continuum image and line cubes for the 146 target clumps have angular resolutions of $\sim 1\farcs2$--$1\farcs9$ (for reference, $2\arcsec$ 
corresponds to 0.1\,pc at a distance of 10\,kpc), and maximum recoverable angular scales $\sim 14\farcs5$--$20\farcs3$. The sensitivity of the 12m-array data is better than 
10\,mJy~beam$^{-1}$ per 0.122\,MHz channel (see Table\,1 of Paper\,I) 
for lines, and $\sim0.4$\,mJy~beam$^{-1}$ for continuum.

\section{Analysis and Results}
\label{sec:result}
\subsection{Core extraction}
\label{sec:gas:distr}
\begin{figure*}
\centering
\includegraphics[width=3.4 in]{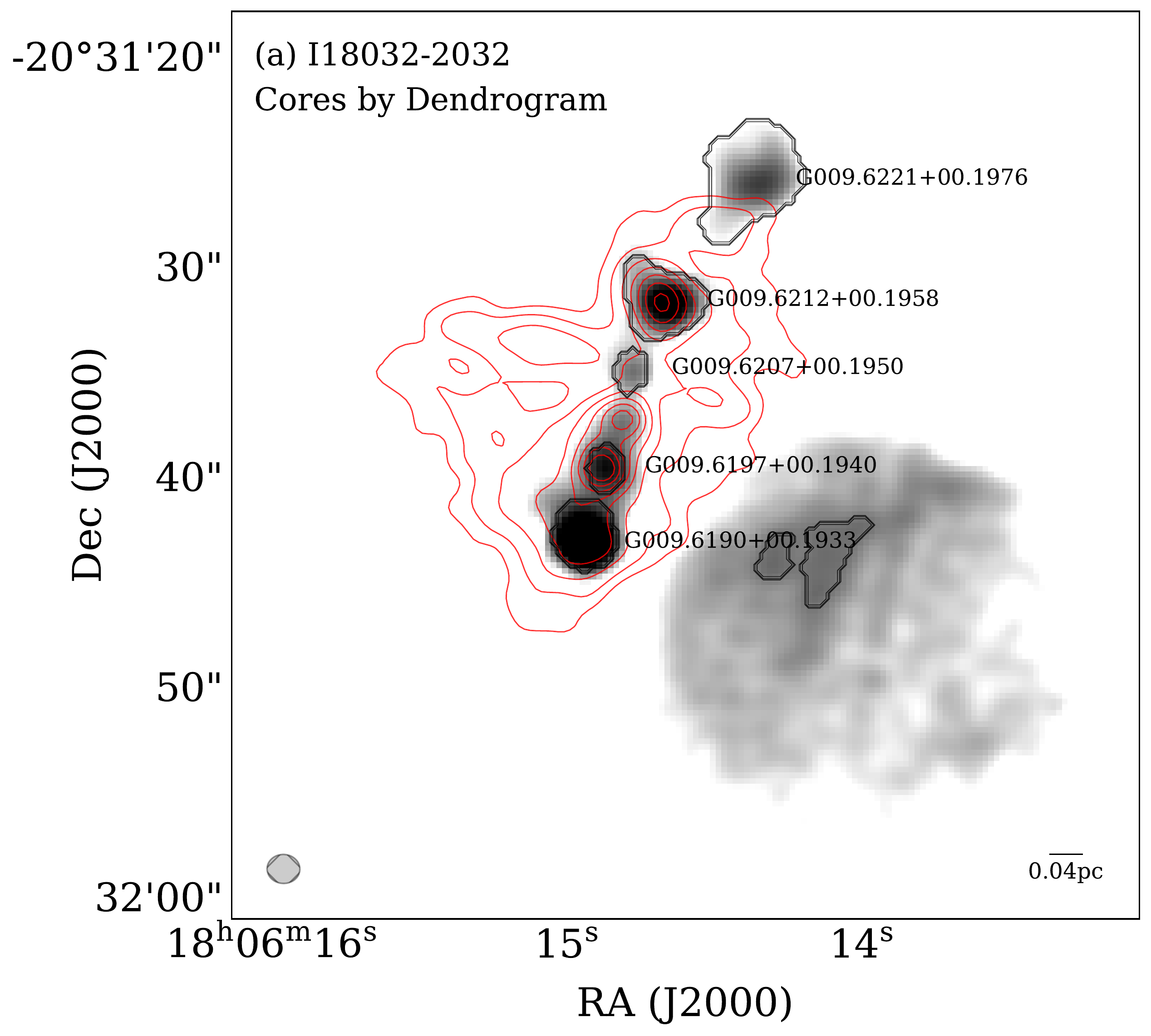}
\includegraphics[width=3.4 in]{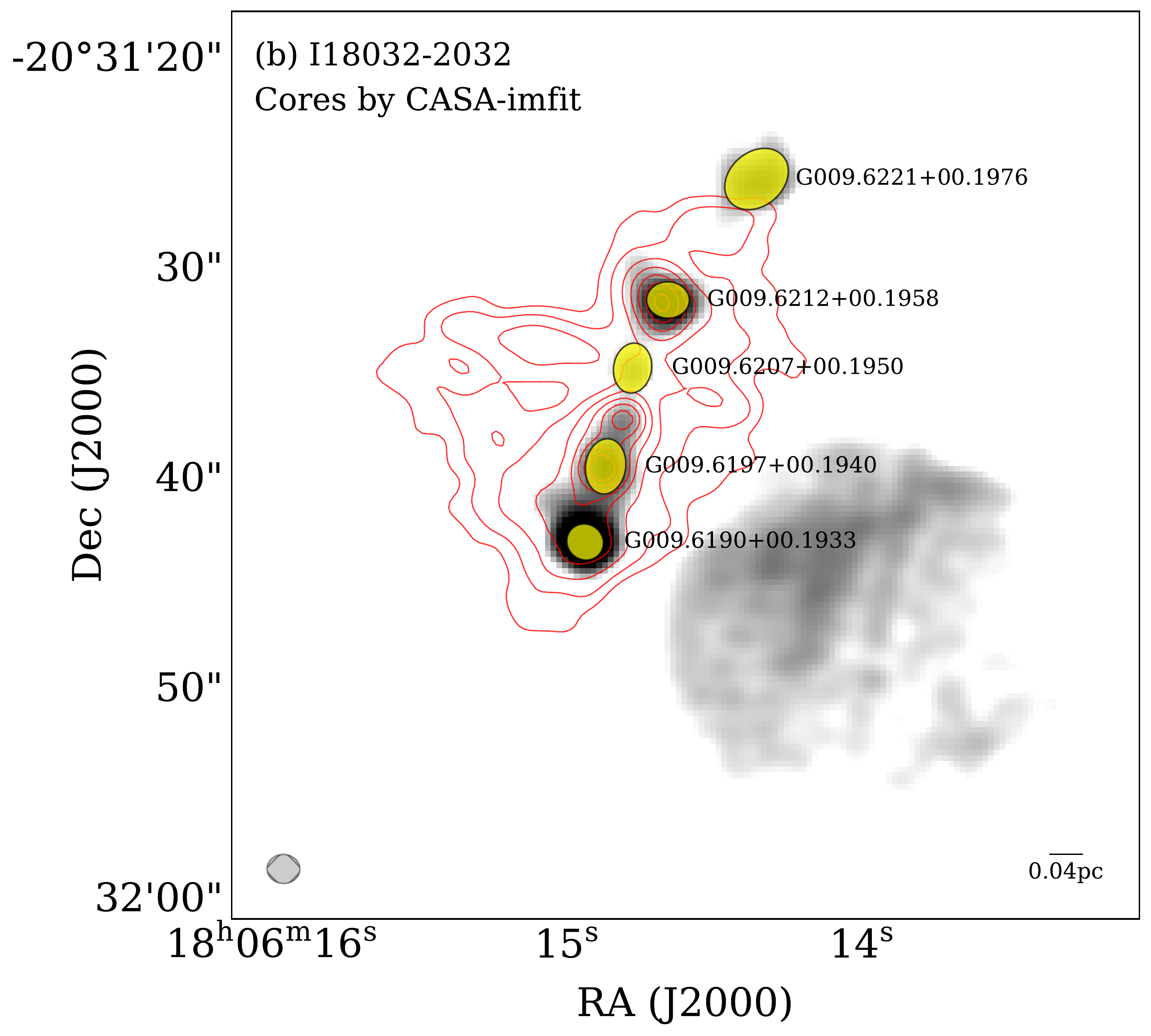}
\caption{Illustration of the core extraction. (a): the preliminary extraction of seven core structures (in black contour) in the 
target clump I18032-2302 from 3\,mm continuum emission by {\it Dendrogram}. (b): the final extraction of 
five finer cores (in yellow ellipses) by CASA-{\it imfit}. In both panels, 3\,mm continuum (in grayscale) is overlaid with  the velocity-integrated
intensity of \hciiin\ (in red contours) over [-35, 45]\,\vel.
The cores labelled with {\it Galactic} names are the  cores remaining after post-selection (see text). One core,  between G009.6197+00.1940 and G009.6207+00.1950, is missing 
in the second extraction procedure due to either insufficient angular resolution or low 
contrast against its background. Note that the contour levels start at 3\,rms with the steps set by a dynamically determined power-law of the form $D=3\times N^{p}+2$, 
where $D$ is the dynamic range of the intensity map (i.e., the ratio between the peak and the rms noise), $N$ is the number of the contours used (8 in this case), 
and $p$ the contour power-law index. The beam of the 3\,mm continuum map and a scale bar of 0.04\,pc are indicated at the bottom left, and bottom right of each panel,
respectively.
}
\label{fig:identify-classify-example}
\end{figure*}

To extract compact cores from the 3\,mm continuum maps,
we make use of both the {\it Dendrogram} algorithm\footnote{\url{https://dendrograms.readthedocs.io/en/stable/}} and CASA--{\it imfit} function.
{\it Dendrogram} can be easily used to extract the dense leaf-like structures (hereafter referred to as cores), which are the smallest structures without substructures in the terminology of the algorithm, and thus customarily taken to be the candidates for cores. 
This technique does not always provide  good measurements of the core parameters such as size and position angle, while 
CASA-{\it imfit} performs better in this regard through a two-dimensional Gaussian fit to the emission.
A visual inspection of the continuum images of the 146 target sources, shows that their outer parts, i.e., the area beyond a radius of 0.01\degr\ from the 
image center, are much noisier than the inner parts, where the 0.01\degr\ radius  corresponds to the pblimit~$=0.4$. Therefore, we masked out the outer part of the images before identifying the core structures. To identify cores we followed two steps. First, using  {\it Dendrogram} we find cores and determine their parameters (i.e., centre position, peak flux density, minor and major-axis sizes, and  position angle), toward each target clump.  Then we made a more accurate measurement of the core parameters  using the CASA task {\it imfit} adopting as initial guesses the parameters determined in the previous step. This approach has been employed in \citet{Li 20}, and demonstrated to work very well. 

In the first step, as inputs of the three key parameters to the {\it Dendrogram} algorithm we used: (i)  
{\it min\_value}~$=2 \sigma_{\rm 3mm}$, where
$\sigma_{\rm 3mm}$ is the noise level. In practice, we first run the {\it Dendrogram} algorithm to identify the  compact cores in the 3\,mm continuum image, then subtract them to obtain a nearly-flat residual 3\,mm image, and
finally estimate the rms level of the residual image as the noise ($\sigma_{\rm 3mm}$) of the original 3\,mm continuum image as the starting level to construct the dendrogram.  In addition, we used
(ii) {\it min\_delta}~$=\sigma_{\rm 3mm}$, the default value in the algorithm to recognise an independent leaf (i.e., the core structure); and (iii) {\it min\_npix}~$=N$ pixels, where $N$ was chosen to be equivalent to the area of one beam (depending on the pixel and beam sizes of the image of each target clump).  This was imposed to ensure a  core to be resolvable. 
Given the above three parameters, the {\it Dendrogram} algorithm automatically locates the cores, and measures
their parameters (see above). To avoid spurious  cores, 
we performed a post-selection to require the selected cores must have peak flux greater than $5 \sigma_{\rm 3mm}$.

In the second step, we used the parameters of the post-selected  cores determined in the first step as inputs to run the CASA--{\it imfit} task  to accurately identify the cores, and measure their parameters, including
peak position, major and minor sizes ($FWHM_{\rm maj}$, $FWHM_{\rm min}$), position angle ($PA$), peak flux ($F_{\rm 3mm}^{\rm p}$),  and integrated flux ($F_{\rm 3mm}^{\rm int}$). 
The measurement uncertainties on the flux are given by 
CASA-imfit. The uncertainties related to the missing flux effect are not taken into account in this work since it is found not significant for compact cores by comparing the 12M data with the combined 12M+7M data (Paper\,I).
Through careful visual inspection of  the 3\,mm continuum map overlaid with the identified  cores, 
we removed a few fake cores with poorly-fitted shapes.
Such cores tend to have a large aspect ratio ($>3$) between the major and minor-axis sizes, appearing either as a filamentary structure
or as a diffuse emission feature. 
As a result, we finally obtained 453 cores from the 146 ATOMS target clumps.

The entire extraction procedure is illustrated in Fig.\,\ref{fig:identify-classify-example},
where we take the target clump I18032-2032 as an example of identifying the cores that are contained. The procedure starts from the  compact core extraction  on the 3 mm continuum map with Dendrogram in the left panel, and then moves to
the finer core fitting with CASA-imfit in the right panel.  Due to the marginal angular resolution ($\sim 2\arcsec$) and small intensity contrast in the 3\,mm continuum emission maps, some cores may be missed in the core extraction.
For example, comparing with the higher intensity-contrast range \hciiin\ (11-10) emission map, one can see that
one molecular core between G009.6197+00.1940 and G009.6207+00.1950 revealed by HC$_3$N (11-10) is missing in the  core extraction procedure. This is because that core cannot be separated from its adjacent cores or background emission in the 3 mm continuum emission map but emerges in higher intensity-contrast HC$_3$N (11-10) emission map.
This issue very rarely occurs in the full sample and predominantly affects low flux (so presumably low-mass) core (specifically, those in the ``unknown core'' category that contains the cores with unknown nature of star formation so far, see Sect.\,\ref{sec:fourth:cat}). Hence, the
analysis in what follows is largely unaffected by this issue.

In addition, with both Dendrogram and CASA-imfit used in the core extraction, several major parameters measured from both methods can be compared.
Actually, we find that the flux and position measurements are consistent in both methods. However, the size measured by CASA-imfit is larger than that in Dendrogram mainly for the low-flux cores (i.e., in the ``unknown cores'' category, see Sect.\,\ref{sec:fourth:cat}). The position angle measurement in CASA-imfit generally matches the 3\,mm continuum dense structure orientation better than that in Dendrogram. These differences mainly come from the different methods of parameter determination used in Dendrogram and CASA-imfit. In the former the size and position angle parameters are derived from the intensity-weighted second moments along the two spatial dimensions \citep{Ros08} while in the latter the two parameters are directly derived from Gaussian fitting to the unweighted intensity spatial distribution. Therefore, to better reflect the core geometry as seen in 3\,mm continuum, we prefer to use the core parameters measured by CASA-imfit.

\subsection{Search for candidate HMCs and \chii\ cores}
\label{sec:core:classify}

One important step towards understanding high-mass star formation is to search for the massive cores that harbor the early stages of high-mass star and cluster formation.  Clues to high-mass star formation can be inferred from the properties of massive starless and/or prestellar cores \egcite{Yua17}, the SED of high-mass protostellar objects (HMPOs, e.g., \citealt{Rat05}), 
HMCs \egcite{Kur00, Rat11}, \chii\ regions \egcite{Chu02, Hoa07}, and some masers (e.g., Class\,II CH$_3$OH masers), which exclusively trace high-mass star formation \egcite{Wan14}.
Here, we focus on the search of massive cores associated with HMCs and \chii\ regions.  The advantage of the ATOMS survey strategy is that the two wide SPWs\,7--8 cover many transitions of the COMs (i.e., HMC tracers; see Fig.\,12 of Paper\,I) and the \halpha\ transition (i.e., the ionized gas tracer).
In what follows, we discuss the  search 
for  \chii\ cores in Sect.\,\ref{sec:core:classify-UCHii} and COM-containing cores in Sect.\,\ref{sec:core:classify-HMC} that could correspond to candidate
HMCs.

\subsubsection{Search for  \chii\ Cores}
\label{sec:core:classify-UCHii}
\begin{figure*}
\centering
\includegraphics[width=6.0 in, height=2.5 in]{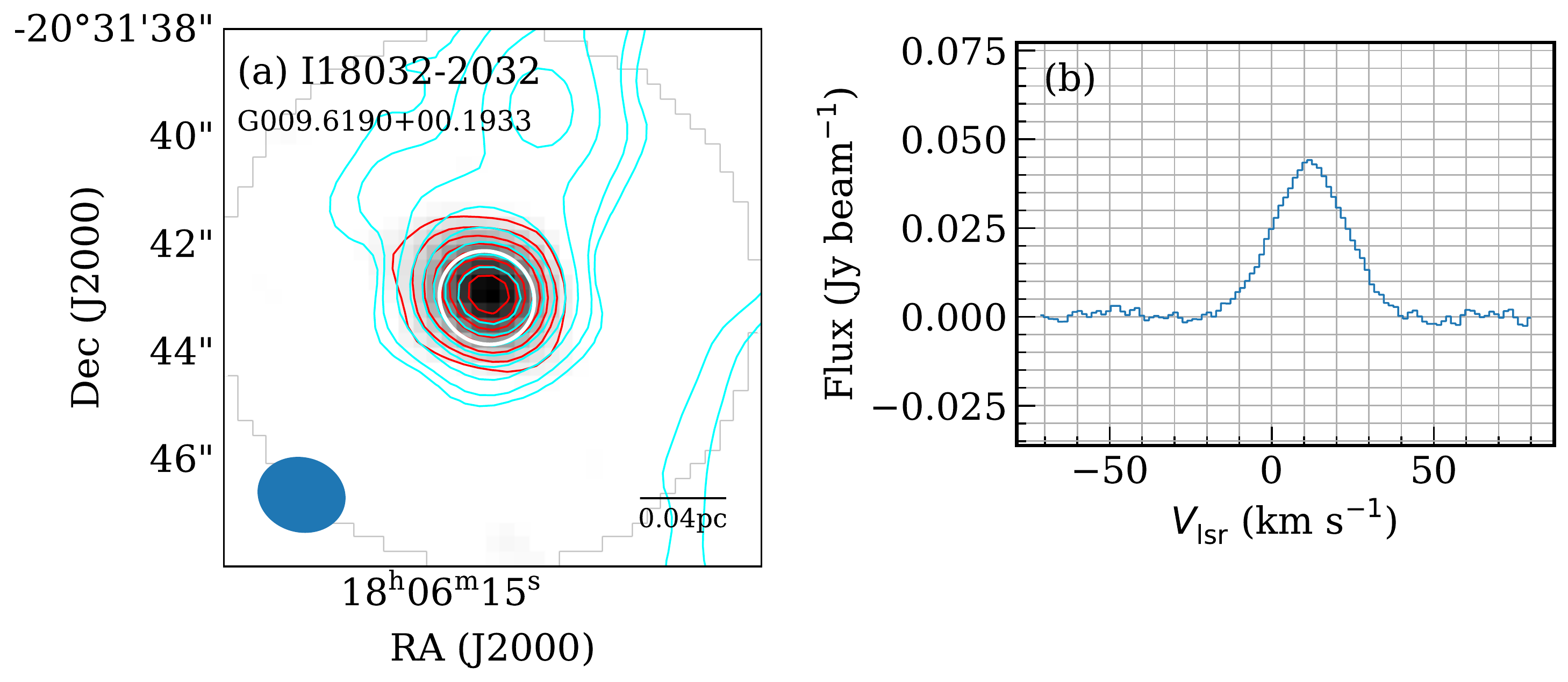}
\caption{Illustration of the \chii\ core search. (a): velocity-integrated intensity map of \halpha\ (in both grayscale and red contours)
 over [-70, 80]\,\vel\ for the core G009.6190+00.1933  (in white ellipse) in  
 the target clump I18032-2032. The map is centered at the core peak with a size of 10\arcsec.
 For comparison, the 3\,mm continuum is also superimposed in cyan contours. The synthesized beam and a scale bar of 0.04\,pc are drawn at the bottom of the panel.
(b): beam ($2\arcsec$)-averaged spectrum of \halpha\ at the core centre. 
G009.6190+00.1933 is an \chii\ core as characterized by strong, and compact \halpha\ emission.
}
\label{fig:uchiicore-classify-i}
\end{figure*}
The entire procedure of the \chii\ core search is illustrated in Fig.\,\ref{fig:uchiicore-classify-i},
where the core G009.6190+00.1933 from the clump I18032-2032 is taken as an example.
In  Fig.\,\ref{fig:uchiicore-classify-i}a, the velocity-integrated intensity of \halpha\ emission over a velocity range of  150\,\vel\ centered at the systemic velocity of the
 core (in both the grayscale and red contours)
is overlaid with the 3\,mm  continuum (cyan contours). The core in question is indicated by the white ellipse.  
Figure\,\ref{fig:uchiicore-classify-i}b presents the beam-averaged spectrum of \halpha\ at the core center.
If the \halpha\ emission appears compact and is spatially associated with the compact continuum core, the core will be classified as an \chii\ core. Following this rule, we finally obtain 
89 \chii\ cores. Their parameters determined by CASA--{\it imfit} are given in Table\,\ref{tbl:uchii}, including 
the clump and core names (i.e., Cols.\,1--2),
the measured sizes of the major and minor axes and the position angle (i.e., Cols.\,4--5), the deconvolved sizes of both axes 
(i.e., Cols.\,6), the core-integrated 3\,mm flux (i.e., Col.\,7), the core-peak flux (i.e., Col.\,8).
The radii (i.e., Col.\,9) of the cores were calculated assuming the same distance as that of their natal clumps (see Paper\,I).
In addition, we measured the line peak intensity ($F_{\rm H40_{\alpha}}^{\rm p}$), the velocity centroid ($V_{\rm lsr, H40_{\alpha}}$), the line width ($\Delta V_{\rm H40_{\alpha}}$) of the \halpha\ line
for each \chii\ core by fitting a single-Gaussian component to the beam-averaged spectrum of \halpha. For comparison (see Sect.\,\ref{sec:disc:size-FWHM-ha}), the deconvoluted size of the \chii\ cores from \halpha\ emission was also measured with CASA-imfit and the corresponding
radius ($\Delta R_{\rm core}^{\rm H40_{\alpha}}$) was computed as above.
These parameters can also be found in last five columns of Table\,\ref{tbl:uchii}.
The $\Delta R_{\rm core}^{\rm H40_{\alpha}}$ and $\Delta V_{\rm H40_{\alpha}}$ together suggest that the cores with associated compact \halpha\ emission are indeed \chii\ sources (see 
Sect.\,\ref{sec:disc:size-FWHM-ha}).

\begin{table*}
\centering
\caption{Parameters of the \chii\ cores from the ATOMS survey's clumps.}
\label{tbl:uchii}
\resizebox{18cm}{!}{
\begin{tabular}{llccccccllcccc}
\hline\hline
           Name &            Gal. Name &             $D$ &          $FWHM$ &              PA & $FWHM_{\rm dec}$ & $F_{\rm 3mm}^{int.}$ & $F_{\rm 3mm}^{p}$ &  $R_{\rm core}$ & $F_{\rm H40\alpha}^{p}$ & $V_{\rm lsr,H40\alpha}$ & $\Delta V_{\rm H40\alpha}$ & $FWHM_{\rm dec}^{\rm H40\alpha}$ & $R_{\rm core}^{\rm H40\alpha}$ \\
         Clump  &                core   &    kpc & ${\rm maj(\arcsec)  \times min(\arcsec)}$ &    deg & ${\rm maj(\arcsec)  \times min(\arcsec)}$ &    mJy & mJy beam$^{-1}$ & 10$^{-2}$ pc & mJy beam$^{-1}$ & km s$^{-1}$ & km s$^{-1}$ & ${\rm maj(\arcsec)  \times min(\arcsec)}$ & 10$^{-2}$ pc \\

\hline
     I09002-4732  &   G268.4225-00.8491  &     2.0  $_{-     0.8  }  ^{+     0.7  }$  &     4.6  $\times$     4.0  &    15.8  &     4.5  $\times$     3.6  &  3058.0  $\pm$    29.0  &   329.7  $\pm$     2.9  &     2.0  $_{-     0.8  }  ^{+     0.6  }$  &   242.7  $\pm$     1.5  &    -2.0  $\pm$     1.6  &    34.5  $\pm$     1.6  &     4.4  $\times$     3.3  &     1.9  $_{-     0.7  }  ^{+     0.6  }$  \\ 
     I12320-6122  &  G300.9688+01.1475*  &     3.1  $_{-     0.5  }  ^{+     0.6  }$  &     2.6  $\times$     2.4  &    11.2  &     1.6  $\times$     0.9  &   477.4  $\pm$     9.5  &   333.5  $\pm$     4.3  &     0.9  $_{-     0.1  }  ^{+     0.2  }$  &   191.6  $\pm$     2.0  &   -41.4  $\pm$     1.6  &    37.3  $\pm$     1.6  &     1.5  $\times$     0.7  &     0.8  $_{-     0.1  }  ^{+     0.2  }$  \\ 
     I12326-6245  &  G301.1364-00.2257*  &     5.6  $_{-     0.8  }  ^{+     0.5  }$  &     2.9  $\times$     2.2  &   172.0  &     1.8  $\times$     1.0  &  2003.0  $\pm$    49.0  &  1351.0  $\pm$    21.0  &     1.8  $_{-     0.3  }  ^{+     0.1  }$  &   526.5  $\pm$     3.2  &   -63.8  $\pm$     1.6  &    55.0  $\pm$     1.6  &     1.6  $\times$     0.8  &     1.6  $_{-     0.2  }  ^{+     0.1  }$  \\ 
     I12383-6128  &   G301.7314+01.1033  &     2.8  $_{-     0.6  }  ^{+     0.7  }$  &     4.1  $\times$     3.7  &    17.0  &     3.3  $\times$     3.2  &    96.7  $\pm$     2.2  &    27.4  $\pm$     0.5  &     2.2  $_{-     0.5  }  ^{+     0.5  }$  &    32.8  $\pm$     0.9  &   -38.3  $\pm$     1.6  &    24.7  $\pm$     1.6  &     3.2  $\times$     2.3  &     1.9  $_{-     0.4  }  ^{+     0.4  }$  \\ 
     I12572-6316  &  G303.9302-00.6879*  &    11.5  $_{-     0.3  }  ^{+     0.3  }$  &     2.8  $\times$     2.3  &     7.3  &     1.5  $\times$     1.3  &    48.2  $\pm$     1.0  &    32.3  $\pm$     0.5  &     3.9  $_{-     0.1  }  ^{+     0.1  }$  &    17.9  $\pm$     0.8  &    24.0  $\pm$     1.6  &    24.5  $\pm$     1.6  &     1.4  $\times$     1.0  &     3.2  $_{-     0.1  }  ^{+     0.1  }$  \\ 
     I13080-6229  &  G305.1961+00.0331*  &     2.3  $_{-     0.6  }  ^{+     0.6  }$  &    12.6  $\times$     8.9  &    78.0  &    12.5  $\times$     8.6  &  2014.0  $\pm$    46.0  &    76.6  $\pm$     1.7  &     5.8  $_{-     1.4  }  ^{+     1.5  }$  &    70.0  $\pm$     1.8  &   -35.8  $\pm$     1.6  &    28.0  $\pm$     1.6  &     9.3  $\times$     5.2  &     3.9  $_{-     1.0  }  ^{+     1.0  }$  \\ 
     I13291-6249  &   G307.5605-00.5870  &     7.8  $_{-     0.5  }  ^{+     0.5  }$  &     7.6  $\times$     5.9  &    24.2  &     7.3  $\times$     5.6  &   562.4  $\pm$     9.3  &    53.1  $\pm$     0.8  &    12.1  $_{-     0.7  }  ^{+     0.8  }$  &    51.0  $\pm$     1.5  &   -38.2  $\pm$     1.6  &    28.6  $\pm$     1.6  &     4.4  $\times$     3.2  &     7.1  $_{-     0.4  }  ^{+     0.4  }$  \\ 
     I13471-6120  &  G309.9206+00.4788*  &     6.6  $_{-     0.6  }  ^{+     0.6  }$  &     2.4  $\times$     1.9  &    71.9  &     1.0  $\times$     1.0  &   848.0  $\pm$    21.0  &   648.0  $\pm$    10.0  &     1.6  $_{-     0.1  }  ^{+     0.1  }$  &   347.8  $\pm$     1.8  &   -57.0  $\pm$     1.6  &    34.1  $\pm$     1.6  &     1.1  $\times$     1.0  &     1.6  $_{-     0.1  }  ^{+     0.1  }$  \\ 
     I14013-6105  &   G311.6273+00.2902  &     3.5  $_{-     0.5  }  ^{+     0.7  }$  &     5.0  $\times$     4.8  &   155.0  &     4.8  $\times$     4.3  &  1198.0  $\pm$    34.0  &   171.0  $\pm$     4.3  &     3.9  $_{-     0.5  }  ^{+     0.8  }$  &   133.1  $\pm$     1.5  &   -58.3  $\pm$     1.6  &    37.6  $\pm$     1.6  &     4.1  $\times$     3.0  &     3.0  $_{-     0.4  }  ^{+     0.6  }$  \\ 
     I14050-6056  &   G312.1084+00.3093  &     3.1  $_{-     0.5  }  ^{+     0.5  }$  &     5.4  $\times$     4.8  &    78.3  &     5.0  $\times$     4.5  &   487.0  $\pm$    16.0  &    63.3  $\pm$     1.9  &     3.5  $_{-     0.6  }  ^{+     0.6  }$  &    50.7  $\pm$     2.0  &   -46.5  $\pm$     1.6  &    41.0  $\pm$     1.6  &     3.5  $\times$     2.6  &     2.2  $_{-     0.4  }  ^{+     0.4  }$  \\ 

\hline
\end{tabular}
}

\begin{flushleft}
{ NOTE:}  The $*$ symbol after the core name indicates that this core is associated with a COM-containing core or COM-containing cores.
$FWHM$ is the measured  core size while
$FWHM_{\rm dec}$ is the corresponding deconvolved size. 
$R_{\rm core}$ follows the form of $R_{\rm eff}/3600\times \pi/180\times D$, where $R_{\rm eff}$ is the effective radius equal to $\sqrt{FWHM_{\rm dec}}/2$, and $D$ is the distance of the core. The errors on $R_{\rm core}$ and $R_{\rm core}^{\rm H40_{\alpha}}$ mainly result from the distance uncertainties and those on the fluxes arise from the 2D Gaussian fitting in the core extraction.
Only a small portion of the data is provided here and the full table will be available as supplementary material.

\end{flushleft}
\end{table*}

\subsubsection{Search for COM-containing cores}
\label{sec:core:classify-HMC}
\begin{figure*}
\centering
\includegraphics[width=6.8 in,  height=2.5 in]{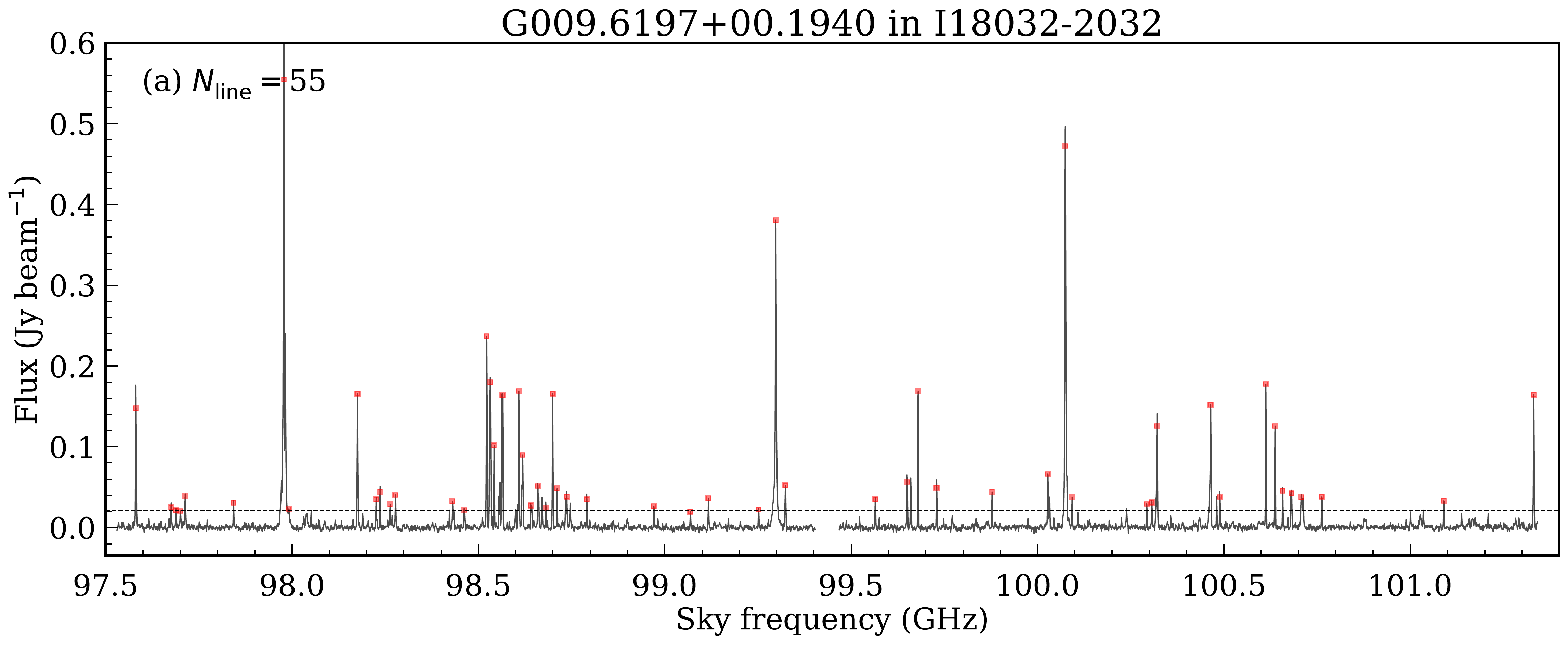}
\includegraphics[width=6.9 in,  height=2.5 in]{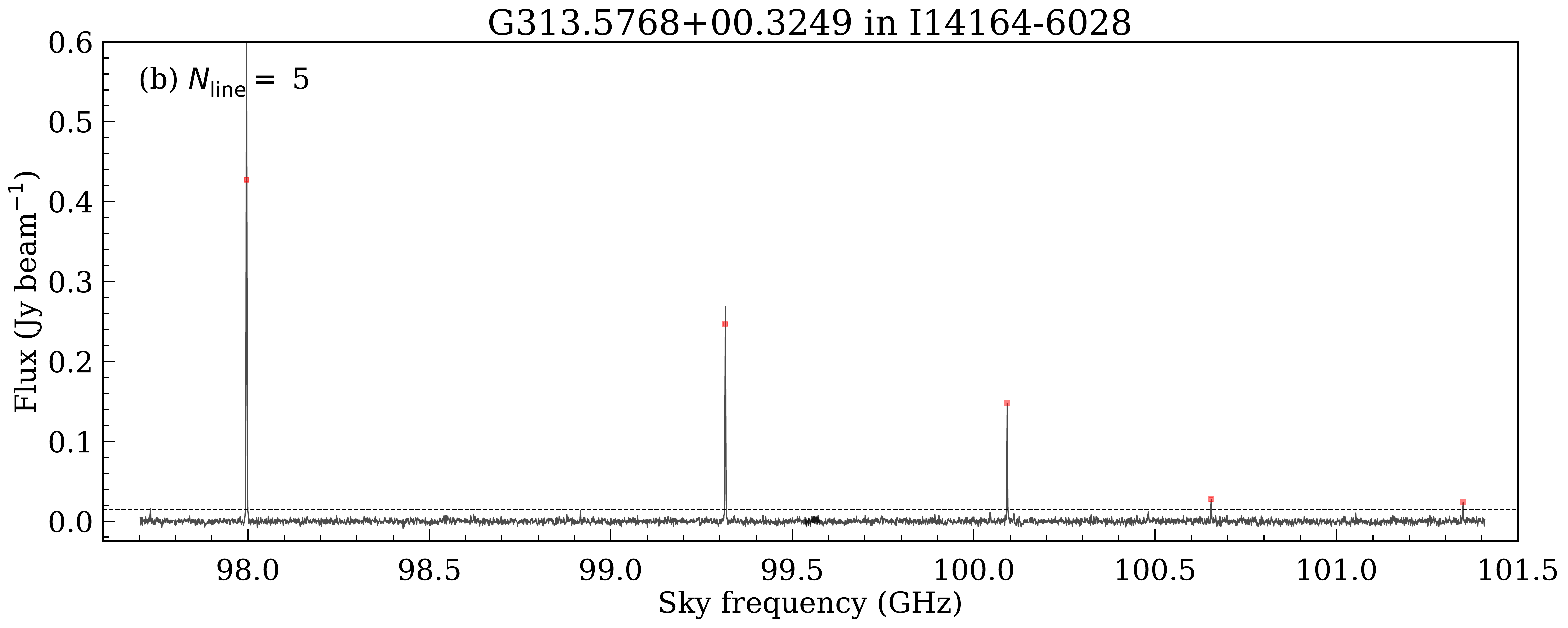}
\caption{ Beam (2\arcsec)-averaged spectrum of SPWs\,7--8 at the emission peak 
of the two COM-containing cores (see definition in text) from two ATOMS clumps. The core and clump names are labelled on top of each panel. 
The red squares indicate the emission lines above 3\,rms  and their total number is given 
to the parameter $N_{\rm line}$. The horizontal dashed line indicates the 3\,rms noise level across the two spectral windows.
For reference, the five lines of panel\,b in an order of decreasing intensity are
CS~(2-1), SO~(3-2), \hciiin~(11-10), \chiiioh\ and H$_2$CO, respectively. 
}
\label{fig:core-sp-wideband}

\end{figure*}

\begin{figure*}
\centering
\includegraphics[width=6.0 in, height=2.5 in]{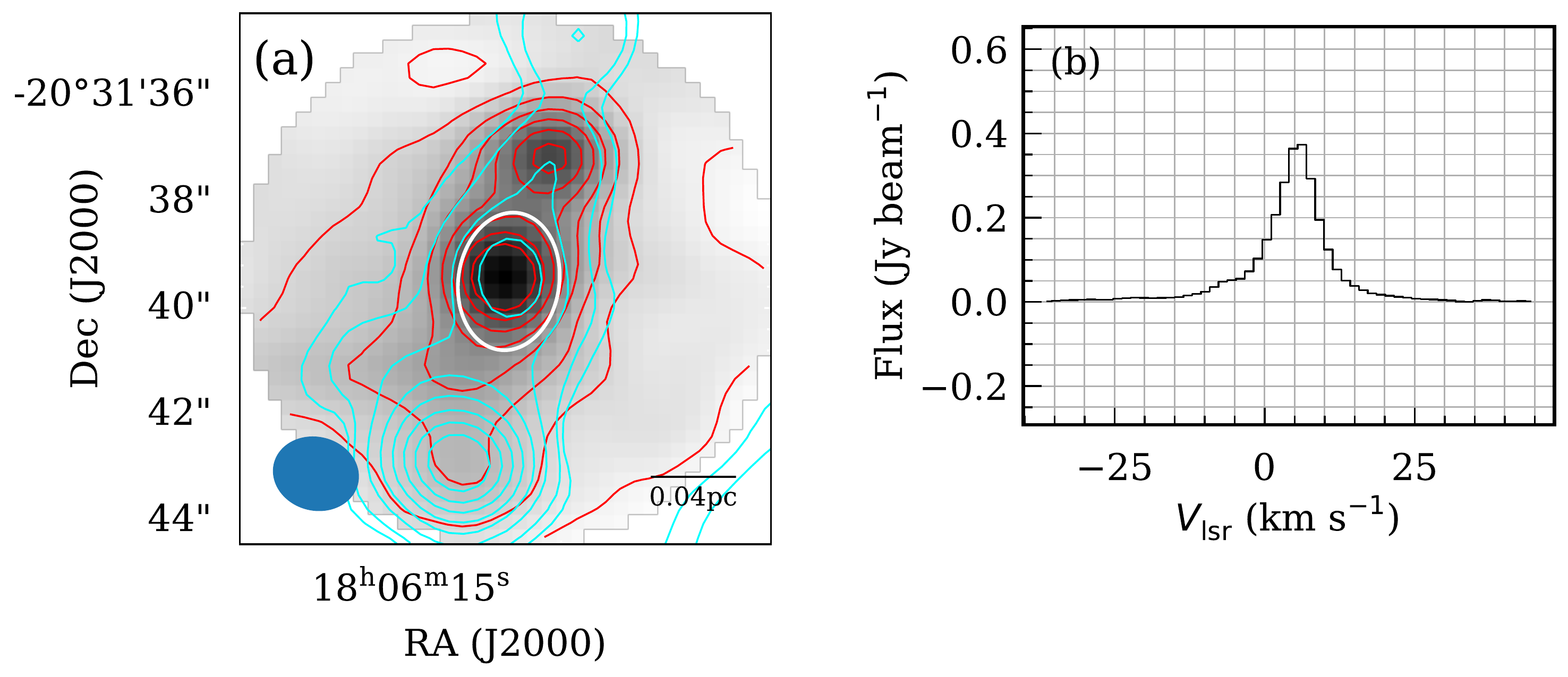}
\caption{ 
(a): velocity-integrated intensity map of \hciiin\ over [-35, 45]\,\vel\ (in both grayscale and red contours) for
 the G009.6197+00.1940 core (in white ellipse) and its surroundings. 
 The map is centred at the core peak with a size of 10\arcsec.
 For comparison, the 3\,mm continuum
 is superimposed in cyan contours. The synthesized beam and a scale bar of 0.04\,pc are drawn at the bottom left of the panel.
(b): beam ($2\arcsec$)-averaged spectrum of \hciiin\ at the core centre (i.e., the white ellipse in panel\,a). 
}
\label{fig:core-classify-i}
\end{figure*}

\begin{figure}
\centering
\includegraphics[width=3.4 in, height=2.0 in]{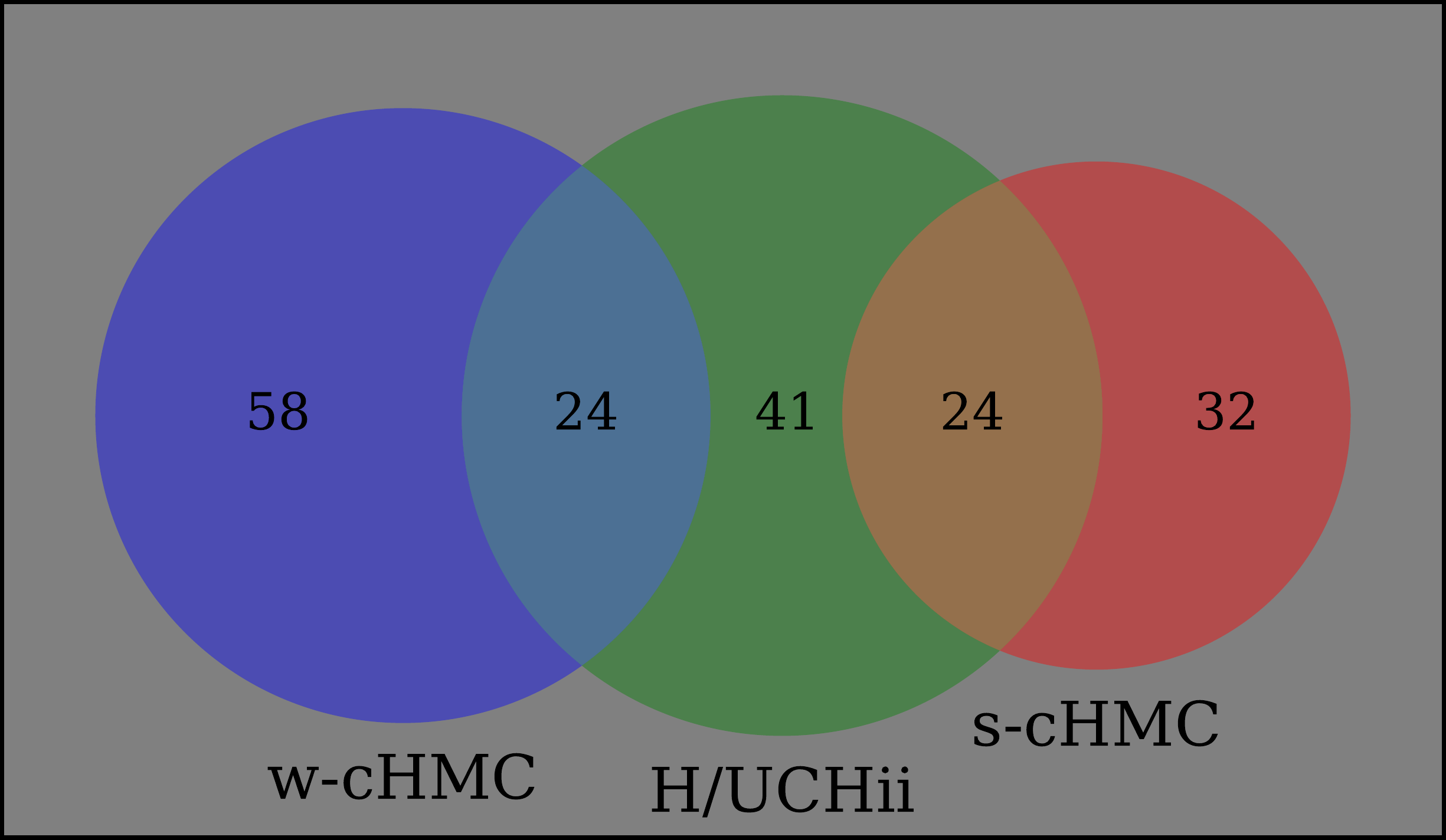}
\caption{Venn diagram showing the relation between 138 COM-containing and 89 \chii\ cores. The 138 COM-containing cores consist of 58 pure and 24 \chii--associated \PCHMC s and 32 pure and 24 \chii--associated \CHMC s.
}
\label{fig:core-venn1}
\end{figure}

\begin{table*}
\centering
\caption{Parameters of the COM-containing  but not \chii-associated cores with  $N_{\rm line}\ge20$ (or  pure \CHMC).}
\label{tbl:hmc-candidate}
\resizebox{18cm}{!}{
\begin{tabular}{llccccccllccccc}
\hline\hline
            Name &            Gal. Name &             $D$ &          $FWHM$ &              PA & $FWHM_{\rm  dec}$ & $F_{\rm 3mm}^{int.}$ & $F_{\rm 3mm}^{p}$ &  $R_{\rm core}$  &  $N_{\rm line}$ & log($M_{\rm core}$) & log($n_{\rm core}$) & $\Sigma_{\rm core}$ \\
           Clump &                 Core &    kpc & ${\rm maj(\arcsec)  \times min(\arcsec)}$ &    deg & ${\rm maj(\arcsec)  \times min(\arcsec)}$ &    mJy & mJy beam$^{-1}$ & 10$^{-2}$ pc &         & $M_{\odot}$ & cm$^{-3}$ & g cm$^{-2}$ \\

\hline
      I08303-4303  &  G261.6444-02.0890  &     2.4  $_{-     0.6  }  ^{+     0.5  }$  &     2.3  $\times$     2.1  &    58.0  &     1.9  $\times$     1.5  &    18.2  $\pm$     0.8  &     7.0  $\pm$     0.2  &     1.0  $_{-     0.2  }  ^{+     0.2  }$  &      28  &     1.1  $_{-     0.7  }  ^{+     0.7  }$  &     6.6  $_{-     3.9  }  ^{+     3.9  }$  &     8.2  \\ 
      I08470-4243  &  G263.2500+00.5137  &     2.4  $_{-     0.6  }  ^{+     0.6  }$  &     2.6  $\times$     2.1  &   147.6  &     2.1  $\times$     1.5  &    38.0  $\pm$     0.9  &    13.5  $\pm$     0.2  &     1.0  $_{-     0.3  }  ^{+     0.2  }$  &      61  &     1.4  $_{-     1.1  }  ^{+     1.1  }$  &     6.8  $_{-     4.2  }  ^{+     4.2  }$  &    15.6  \\ 
      I13134-6242  &  G305.7992-00.2447  &     2.0  $_{-     0.4  }  ^{+     0.6  }$  &     3.4  $\times$     2.7  &   171.5  &     2.4  $\times$     2.1  &   103.8  $\pm$     2.4  &    47.2  $\pm$     0.8  &     1.1  $_{-     0.2  }  ^{+     0.3  }$  &      48  &     1.7  $_{-     1.3  }  ^{+     1.5  }$  &     7.0  $_{-     4.1  }  ^{+     4.2  }$  &    26.6  \\ 
      I13484-6100  &  G310.1436+00.7598  &     6.9  $_{-     0.7  }  ^{+     0.5  }$  &     3.5  $\times$     3.2  &    77.0  &     2.8  $\times$     2.7  &    41.1  $\pm$     1.7  &    13.0  $\pm$     0.4  &     4.6  $_{-     0.5  }  ^{+     0.3  }$  &      36  &     2.4  $_{-     1.7  }  ^{+     1.5  }$  &     5.9  $_{-     4.0  }  ^{+     3.9  }$  &     7.0  \\ 
      I14498-5856  &  G318.0482+00.0863  &     2.9  $_{-     0.3  }  ^{+     0.4  }$  &     5.2  $\times$     2.6  &    48.3  &     4.0  $\times$     2.0  &     9.1  $\pm$     0.0  &     4.2  $\pm$     0.2  &     2.0  $_{-     0.2  }  ^{+     0.3  }$  &      33  &     0.9  $_{-     0.3  }  ^{+     0.4  }$  &     5.8  $_{-     3.1  }  ^{+     3.1  }$  &     1.5  \\ 
      I14498-5856  &  G318.0497+00.0868  &     2.9  $_{-     0.3  }  ^{+     0.4  }$  &     3.9  $\times$     3.0  &    63.4  &     3.5  $\times$     2.4  &    60.7  $\pm$     1.4  &    15.9  $\pm$     0.3  &     2.0  $_{-     0.2  }  ^{+     0.3  }$  &      39  &     1.8  $_{-     1.1  }  ^{+     1.2  }$  &     6.4  $_{-     3.6  }  ^{+     3.7  }$  &     9.3  \\ 
      I15394-5358  &  G326.4747+00.7030  &     2.4  $_{-     0.2  }  ^{+     0.3  }$  &     5.2  $\times$     4.4  &   140.9  &     4.9  $\times$     4.0  &   226.7  $\pm$     5.4  &    28.4  $\pm$     0.6  &     2.5  $_{-     0.2  }  ^{+     0.3  }$  &      33  &     2.2  $_{-     1.4  }  ^{+     1.6  }$  &     6.6  $_{-     3.5  }  ^{+     3.6  }$  &    14.9  \\ 

\hline
\end{tabular}
}

\begin{flushleft}
{ NOTE:}  The parameters $FWHM$, $FWHM_{\rm dec}$, and $R_{\rm core}$ are derived in the same way as in Table\,\ref{tbl:uchii}. The errors on $R_{\rm core}$, $M_{\rm core}$ and $n_{\rm core}$ mainly result from the distance uncertainties, and those on the fluxes from the 2D Gaussian fitting in the core extraction. 
Only a small portion of the data is provided here and the full table will be available as supplementary material.
\end{flushleft}
\end{table*}

\begin{table*}
\centering
\caption{Parameters of the COM-containing  but not \chii-associated cores with  $5\le N_{\rm line}<20$ (or pure \PCHMC).}
\label{tbl:hmc-potential}
\resizebox{18cm}{!}{
\begin{tabular}{llccccccllccccc}
\hline\hline
            Name &            Gal. Name &             $D$  &          $FWHM$ &              PA & $FWHM_{\rm dec}$ & $F_{\rm 3mm}^{int.}$ & $F_{\rm 3mm}^{p}$ &  $R_{\rm core}$  &  $N_{\rm line}$ & log($M_{\rm core}$) & log($n_{\rm core}$) & $\Sigma_{\rm core}$ \\
           Clump &                 Core &    kpc  & ${\rm maj(\arcsec)  \times min(\arcsec)}$ &    deg & ${\rm maj(\arcsec)  \times min(\arcsec)}$ &    mJy & mJy beam$^{-1}$ & 10$^{-2}$ pc &         & $M_{\odot}$ & cm$^{-3}$ & g cm$^{-2}$ \\

\hline
      I08448-4343  &  G263.7737-00.4326  &     1.4  $_{-     0.8  }  ^{+     0.6  }$  &     3.5  $\times$     2.3  &   113.0  &     3.1  $\times$     2.0  &     5.6  $\pm$     1.2  &     1.3  $\pm$     0.2  &     0.9  $_{-     0.5  }  ^{+     0.4  }$  &       9  &     0.1  $_{-     0.1  }  ^{+     0.1  }$  &     5.9  $_{-     3.0  }  ^{+     2.9  }$  &     1.2  \\ 
      I09018-4816  &  G269.1523-01.1279  &     2.8  $_{-     0.7  }  ^{+     0.6  }$  &     4.3  $\times$     2.7  &   178.5  &     4.1  $\times$     2.2  &    51.9  $\pm$     2.1  &     8.6  $\pm$     0.3  &     2.1  $_{-     0.5  }  ^{+     0.4  }$  &      12  &     1.7  $_{-     1.4  }  ^{+     1.3  }$  &     6.3  $_{-     3.8  }  ^{+     3.8  }$  &     7.4  \\ 
      I10365-5803  &  G286.2074+00.1698  &     3.4  $_{-     0.6  }  ^{+     0.6  }$  &     3.6  $\times$     3.0  &    81.9  &     3.1  $\times$     2.5  &    31.7  $\pm$     1.0  &     9.2  $\pm$     0.2  &     2.3  $_{-     0.4  }  ^{+     0.4  }$  &      10  &     1.6  $_{-     1.2  }  ^{+     1.1  }$  &     6.1  $_{-     3.7  }  ^{+     3.6  }$  &     5.3  \\ 
      I11298-6155  &  G293.8274-00.7461  &     9.4  $_{-     0.3  }  ^{+     0.4  }$  &     4.0  $\times$     3.0  &    92.8  &     3.6  $\times$     2.3  &    41.8  $\pm$     1.9  &    11.8  $\pm$     0.4  &     6.6  $_{-     0.2  }  ^{+     0.3  }$  &      10  &     2.6  $_{-     1.5  }  ^{+     1.6  }$  &     5.7  $_{-     3.7  }  ^{+     3.8  }$  &     6.5  \\ 
      I11332-6258  &  G294.5115-01.6213  &     1.1  $_{-     0.6  }  ^{+     0.7  }$  &     5.9  $\times$     4.7  &    81.2  &     5.6  $\times$     4.2  &    30.5  $\pm$     2.0  &     3.8  $\pm$     0.2  &     1.3  $_{-     0.7  }  ^{+     0.8  }$  &       6  &     0.6  $_{-     0.7  }  ^{+     0.7  }$  &     5.9  $_{-     2.8  }  ^{+     2.8  }$  &     1.7  \\ 
      I12326-6245  &  G301.1363-00.2235  &     5.6  $_{-     0.8  }  ^{+     0.5  }$  &     3.4  $\times$     2.1  &   178.0  &     2.3  $\times$     1.1  &   212.0  $\pm$    52.0  &   130.0  $\pm$    21.0  &     2.2  $_{-     0.3  }  ^{+     0.2  }$  &       7  &     2.9  $_{-     2.3  }  ^{+     2.1  }$  &     7.1  $_{-     5.2  }  ^{+     4.9  }$  &   108.1  \\ 
      I13079-6218  &  G305.2084+00.2060  &     3.0  $_{-     0.7  }  ^{+     0.6  }$  &     3.5  $\times$     3.0  &    22.8  &     2.8  $\times$     2.2  &    97.0  $\pm$     2.6  &    39.1  $\pm$     0.8  &     1.8  $_{-     0.4  }  ^{+     0.4  }$  &      11  &     2.0  $_{-     1.7  }  ^{+     1.6  }$  &     6.7  $_{-     4.3  }  ^{+     4.2  }$  &    20.3  \\ 
      I13079-6218  &  G305.2022+00.2072  &     3.0  $_{-     0.7  }  ^{+     0.6  }$  &     3.5  $\times$     3.2  &   105.0  &     3.0  $\times$     2.0  &    20.5  $\pm$     2.8  &     7.8  $\pm$     0.8  &     1.8  $_{-     0.4  }  ^{+     0.4  }$  &      11  &     1.3  $_{-     1.0  }  ^{+     0.9  }$  &     6.0  $_{-     3.6  }  ^{+     3.5  }$  &     4.4  \\ 

\hline
\end{tabular}
}

\begin{flushleft}
{ NOTE:} The parameters $FWHM$, $FWHM_{\rm dec}$, and $R_{\rm core}$ are derived in the same way as in Table\,\ref{tbl:uchii}. The errors on $R_{\rm core}$, $M_{\rm core}$ and $n_{\rm core}$ mainly result from the distance uncertainties, and those on the fluxes from the 2D Gaussian fitting in the core extraction. 
Only a small portion of the data is provided here and the full table will be available as supplementary material.
\end{flushleft}
\end{table*}

To quantitatively describe the chemical richness of a core, we use the number of lines ($N_{\rm line}$) detected in the two line-scan windows SPWs\,7--8. 
These spectral windows cover tens of transitions from COMs (e.g., \chthocho, \chthcho, ${\rm CH_3OH}$, ${\rm C_2H_5CN}$) that
are usually detectable in a HMC, 
and only two or three transitions
of non HMC-tracer molecules (i.e., CS~(2--1), SO~(3--2) or ${\rm H_2CO}$, see Fig.\,12 of Paper\,I). Counting the emission lines is therefore useful in a search for the COM-containing cores.

The number of emission lines was determined by searching 
for emission peaks above the 3\,$\sigma$ level across the entire spectrum of the two SPWs, which was also adopted in \citet{San17}. 
In practice, the 1\,$\sigma$ level was calculated individually for each source from the standard deviation of the amplitudes in the line-free emission channels usually located on either or both ends of the wide SPWs. This method works very well for the sources without a forest of detectable lines. However, it is generally conservative for the sources (e.g., the one in Fig.\,3a) with a forest of detectable lines since it is difficult to find completely line-free channels, leading to an overestimate of the sigma level. 
To avoid  counting spurious peaks, for example from the broad line wings of 
the spectrum of the outflow sources, the separation between the nearest neighbours was required to be at least five velocity channels (i.e., $\sim 8$\,\vel). In addition, a valid peak detection was required to have at least three channels above the 3\,$\sigma$ level.

Figure\,\ref{fig:core-sp-wideband} displays an example of the search for the emission lines from the beam-averaged spectrum of the two SPWs at the peak position of two cores, demonstrating that our method works well in counting emission lines.
Note that the two sources given in the example were in different ALMA observing clusters with the center frequencies slightly different, so that the frequency ranges of the sources are different accordingly.
The final COM-containing cores are defined by the following two criteria, $N_{\rm line}\ge5$ and a compact feature of  \hciiin\ emission.
The former ensures at least one transition of the COMs for each candidate (see above, and Fig.\,\ref{fig:core-sp-wideband}), while the latter requires each candidate to be a  compact source in molecular line emission. Figure\,\,\ref{fig:core-classify-i} presents an example of the compact COM-containing core  in \hciiin\ emission.  Figure\,\,\ref{fig:core-classify-i}a shows the velocity-integrated intensity of \hciiin\ (in both grayscale and red contours) 
over a velocity extent of 80\,\vel\ centered at the systemic velocity of the core (in white ellipse) and 3\,mm continuum (in cyan contours), 
while Fig.\,\,\ref{fig:core-classify-i}b displays the beam-averaged spectrum of \hciiin\ at the core center.

Following the above two criteria, 138 COM-containing cores were selected. We find that 48 out of these 138 COM-containing cores are spatially
associated with \chii\ cores  (positions matching  within a radius of 2\arcsec, comparable to the beam size),
 and thus are \chii-associated candidate HMCs (cHMCs, hereafter). Note that the HMC is a definition from 
the molecular chemistry study, and has been observationally found to overlap the \chii\ regions.
 Given their masses and densities (see Sect.\,\ref{sec:disc_COMnature})，
most of the remaining 90 COM-containing cores are cHMCs 
and they all are thus treated as  HMC candidates for simplicity.

 To make a robust sample of cHMCs, the 138 COM-containing cores are divided into two groups of cHMCs 
using the values of $N_{\rm line}$. As shown in Fig.\,\ref{fig:core-venn1}, one group, regarded as strong-cHMCs (s-cHMCs),
contains 56 high-level COM-containing cores with $N_{\rm line}\ge20$, and the other group, regarded as weak-cHMCs (w-cHMCs), contains 82 low-level COM-containing cores  with $5\le N_{\rm line}<20$. In each group, there are 24 cHMCs associated with \chii\ cores. 
The threshold $N_{\rm line}=20$ is somewhat arbitrary, based on our 
visual inspection of the spectra of all cHMCs, but does not matter too much in the following statistical analysis.
Moreover, to distinguish the cHMCs between with \chii\ and without \chii\ cores, 32 pure \CHMC s and 58 pure \PCHMC s are separated from the entire sample of the 138 cHMCs (see Fig.\,\ref{fig:core-venn1}), and compiled into the two catalogues, one called pure \CHMC s and 
the other one pure \PCHMC s. That is, pure cores refer to those without \chii\ signatures.
The parameters of the two catalogues are tabulated
in Table\,\ref{tbl:hmc-candidate} for pure s-cHMCs, and in Table\,\ref{tbl:hmc-potential}
for pure w-cHMCs, including 
the clump and core names (Cols.\,1--2), the measured sizes of the major and minor axes and the position angle (Cols.\,4--5), the deconvolved sizes of both axes  (Cols.\,6), the core-integrated 3\,mm flux (Col.\,7), the core-peak flux (Col.\,8), the radius (Col.\,9), the number of lines (Col.\,10), 
the mass (Col.\,11), the number density (Col.\,12), and the mass surface density (Col.\,13). The last three parameters were calculated following the equations in Appendix\,\ref{sec:app-mass} for a fixed temperature of 100\,K typical of a hot core.
The median radius is $1.6\times10^{-2}$\,pc in an 
interquartile range (IQR)\footnote{The interquartile range  is a measure of statistical dispersion, being equal to the difference
between 75th and 25th percentiles, or between upper and lower quartiles.} 
of [1.0, 2.3]$\times10^{-2}$\,pc for the catalogue of 
32 pure \CHMC s, and $2.6\times10^{-2}$\,pc in an IQR of [1.7, 5.2]$\times10^{-2}$\,pc for the catalogue of 
58 pure \PCHMC s. These results suggest that most of the 90 COM-containing cores are indeed compact.

Note that contamination from free-free radiation to continuum may not be trivial for the cores associated with \chii. In practice, we calculated at a temperature of 100\,K the masses for two types of cores , pure cHMCs and cHMCs associated with \chii\ sources (i.e., compact H40a emission). In comparison, we found that the mass of the latter cores ($\sim 10^4$\,\msun) is on average about two magnitude higher than that of the former cores ($\sim 10^2$\,\msun). This result suggests that contamination by free-free radiation is non-negligible in calculating the flux-related parameters.
Since this contamination cannot be accurately determined, we do not give the estimate of the continuum flux-related parameters for the \chii\ cores such as the mass.

\subsection{ Compact dense core list not associated with either cHMC or \chii.}
\label{sec:fourth:cat}

\begin{table*}
\centering
\caption{Parameters of  the ``unknown'' cores (i.e., no signature of \chii\ nor cHMC)}.
\label{tbl:unknown-core}
\resizebox{18cm}{!}{
\begin{tabular}{llccccccllccccc}
\hline\hline
            Name &            Gal. Name  &             $D$ &          $FWHM$  &              PA & $FWHM_{\rm dec}$ & $F_{\rm 3mm}^{int.}$ & $F_{\rm 3mm}^{p}$ &  $R_{\rm core}$  & log($M_{\rm core}$) & log($n_{\rm core}$) & $\Sigma_{\rm core}$ \\
          Clump  &                Core  &    kpc &  ${\rm maj(\arcsec)  \times min(\arcsec)}$  &    deg & ${\rm maj(\arcsec)  \times min(\arcsec)}$ &    mJy & mJy beam$^{-1}$ & 10$^{-2}$ pc & $M_{\odot}$  & cm$^{-3}$ &g cm$^{-2}$ \\

\hline
      I08076-3556  &  G253.2971-01.6156  &     1.0  $_{-     0.5  }  ^{+     0.5  }$  &     5.1  $\times$     3.3  &    86.4  &     4.9  $\times$     3.0  &    11.6  $\pm$     1.2  &     1.3  $\pm$     0.1  &     0.9  $_{-     0.5  }  ^{+     0.4  }$  &     0.1  $_{-     0.1  }  ^{+     0.1  }$  &     5.9  $_{-     2.6  }  ^{+     2.5  }$  &     0.9  \\ 
      I08076-3556  &  G253.2930-01.6113  &     1.0  $_{-     0.5  }  ^{+     0.5  }$  &     2.2  $\times$     1.6  &    99.3  &     1.6  $\times$     1.1  &    18.8  $\pm$     0.5  &     9.8  $\pm$     0.2  &     0.3  $_{-     0.2  }  ^{+     0.1  }$  &     0.3  $_{-     0.3  }  ^{+     0.3  }$  &     7.2  $_{-     4.0  }  ^{+     3.9  }$  &    13.0  \\ 
      I08303-4303  &  G261.6446-02.0899  &     2.4  $_{-     0.6  }  ^{+     0.5  }$  &     4.9  $\times$     2.5  &     6.9  &     4.7  $\times$     2.0  &    23.7  $\pm$     1.4  &     3.6  $\pm$     0.2  &     1.8  $_{-     0.4  }  ^{+     0.4  }$  &     1.1  $_{-     0.8  }  ^{+     0.8  }$  &     6.0  $_{-     3.4  }  ^{+     3.3  }$  &     2.9  \\ 
      I08303-4303  &  G261.6444-02.0876  &     2.4  $_{-     0.6  }  ^{+     0.5  }$  &     4.0  $\times$     3.8  &    40.0  &     3.8  $\times$     3.5  &    15.2  $\pm$     1.6  &     1.8  $\pm$     0.2  &     2.1  $_{-     0.5  }  ^{+     0.5  }$  &     1.0  $_{-     0.6  }  ^{+     0.6  }$  &     5.7  $_{-     3.0  }  ^{+     3.0  }$  &     1.3  \\ 
      I08448-4343  &  G263.7776-00.4332  &     1.4  $_{-     0.8  }  ^{+     0.6  }$  &     5.2  $\times$     3.0  &   116.0  &     5.0  $\times$     2.8  &     8.8  $\pm$     1.7  &     1.1  $\pm$     0.2  &     1.3  $_{-     0.7  }  ^{+     0.6  }$  &     0.3  $_{-     0.3  }  ^{+     0.2  }$  &     5.6  $_{-     2.8  }  ^{+     2.7  }$  &     0.7  \\ 
      I08448-4343  &  G263.7729-00.4364  &     1.4  $_{-     0.8  }  ^{+     0.6  }$  &     2.5  $\times$     1.8  &    72.0  &     2.1  $\times$     1.1  &     2.1  $\pm$     0.8  &     0.9  $\pm$     0.2  &     0.5  $_{-     0.3  }  ^{+     0.2  }$  &    -0.3  $_{-    -0.3  }  ^{+    -0.4  }$  &     6.0  $_{-     3.1  }  ^{+     3.0  }$  &     1.1  \\ 
      I08448-4343  &  G263.7724-00.4366  &     1.4  $_{-     0.8  }  ^{+     0.6  }$  &     2.2  $\times$     1.8  &   103.0  &     1.5  $\times$     1.3  &     1.9  $\pm$     0.7  &     0.9  $\pm$     0.2  &     0.5  $_{-     0.3  }  ^{+     0.2  }$  &    -0.4  $_{-    -0.4  }  ^{+    -0.5  }$  &     6.0  $_{-     3.1  }  ^{+     3.0  }$  &     1.1  \\ 
      I08448-4343  &  G263.7723-00.4350  &     1.4  $_{-     0.8  }  ^{+     0.6  }$  &     4.4  $\times$     2.2  &    80.6  &     3.6  $\times$     1.8  &     7.3  $\pm$     0.0  &     5.8  $\pm$     0.2  &     0.9  $_{-     0.5  }  ^{+     0.4  }$  &     0.2  $_{-     0.2  }  ^{+     0.1  }$  &     6.5  $_{-     3.7  }  ^{+     3.6  }$  &     1.3  \\ 
      I08448-4343  &  G263.7744-00.4328  &     1.4  $_{-     0.8  }  ^{+     0.6  }$  &     2.5  $\times$     2.2  &    18.0  &     2.1  $\times$     1.6  &    11.5  $\pm$     0.9  &     3.9  $\pm$     0.2  &     0.6  $_{-     0.3  }  ^{+     0.3  }$  &     0.4  $_{-     0.4  }  ^{+     0.3  }$  &     6.5  $_{-     3.6  }  ^{+     3.5  }$  &     3.9  \\ 
      I08448-4343  &  G263.7712-00.4363  &     1.4  $_{-     0.8  }  ^{+     0.6  }$  &     3.2  $\times$     1.8  &   157.0  &     2.9  $\times$     1.1  &     2.4  $\pm$     0.9  &     0.8  $\pm$     0.2  &     0.6  $_{-     0.3  }  ^{+     0.3  }$  &    -0.3  $_{-    -0.3  }  ^{+    -0.4  }$  &     5.8  $_{-     2.9  }  ^{+     2.9  }$  &     0.9  \\ 

\hline
\end{tabular}
}

\begin{flushleft}
{ NOTE:}  The parameters $FWHM$, $FWHM_{\rm dec}$, and $R_{\rm core}$ are derived in the same way as in Table\,\ref{tbl:uchii}. The errors on $R_{\rm core}$, $M_{\rm core}$ and $n_{\rm core}$ mainly come from the distance uncertainties, those on the fluxes from the 2D Gaussian fitting in the core extraction. 
Only a small portion of the data is provided here, and the full table will be available as supplementary material.

\end{flushleft}
\end{table*}

As described above, we extract 453 cores in 3\,mm continuum.  Of
these, 89 are classified as H/UC-HII, 32 as s-cHMC, and 58 as
w-cHMC.  The remaining 274 cores lack observationally-quantifiable
metrics for association into any of the three categories above.
Hence we refer to these as "unknown cores" (see Table\,\ref{tbl:unknown-core}).  These
``unknown cores" could be simple continuum flux condensations in an
earlier evolutionary stage. Indeed, chemically poor HMPOs and candidate massive prestellar cores have been detected
in some massive proto-clusters \citep[e.g.,][]{Liu17}. However, it is also possible that some (to
date) unknown but potentially significant fraction of these ``unknown cores" may
actually be cHMC or \chii\ cores that fall below our sensitivity
limit due to spatial resolution, distance, and/or flux limitations. 
These will be investigated in more detail in future work. 

For reference, the mass, number density, and mass surface density of 
the ``unknown'' cores (see Table\,\ref{tbl:unknown-core}), were calculated (see Appendix\,\ref{sec:app-mass}) assuming a dust 
temperature of 25\,K. This temperature is comparable to 
the median value of the clumps that do not have cHMC or
\chii\ signatures.
As a result, these cores have a median radius of $2.4\times10^{-2}$\,pc in an IQR of [1.3, 4.6]$\times10^{-2}$\,pc, 
 a median mass of 19\,\msun\ in an IQR of [5, 99]\,\msun, a median number density of 
 $0.6\times10^{6}$\,\pcmcu\ in an IQR of [0.3, 1.3]$\times10^{6}$\,\pcmcu, and a median mass surface density 
 of 2.2\,g~\cmcm\ in an IQR of [1.1, 4.3]\,g~\cmcm.  The properties of the "unknown" cores will be more thoroughly investigated in a forthcoming paper.

\section{Discussion}
\label{sec:disc}

\subsection{Limitations in dense core extraction and classification}
\label{sec:limitations}

Our ATOMS survey consists of a uniform sensitivity and angular
resolution survey toward regions located over a large range of distances.  Consequently, 
the dense core extraction, as discussed above, is affected by
variations of factors of order of $\sim$ 10 in distance across the
sample.  Here we focus on the most robust (brightest) portion of our
sample (i.e. \PCHMC, \CHMC\ and H/UC-H{\sc ii} cores, see Sect.\,\ref{sec:core:classify}). 

\begin{figure*}
\centering
\includegraphics[width=3.4 in]{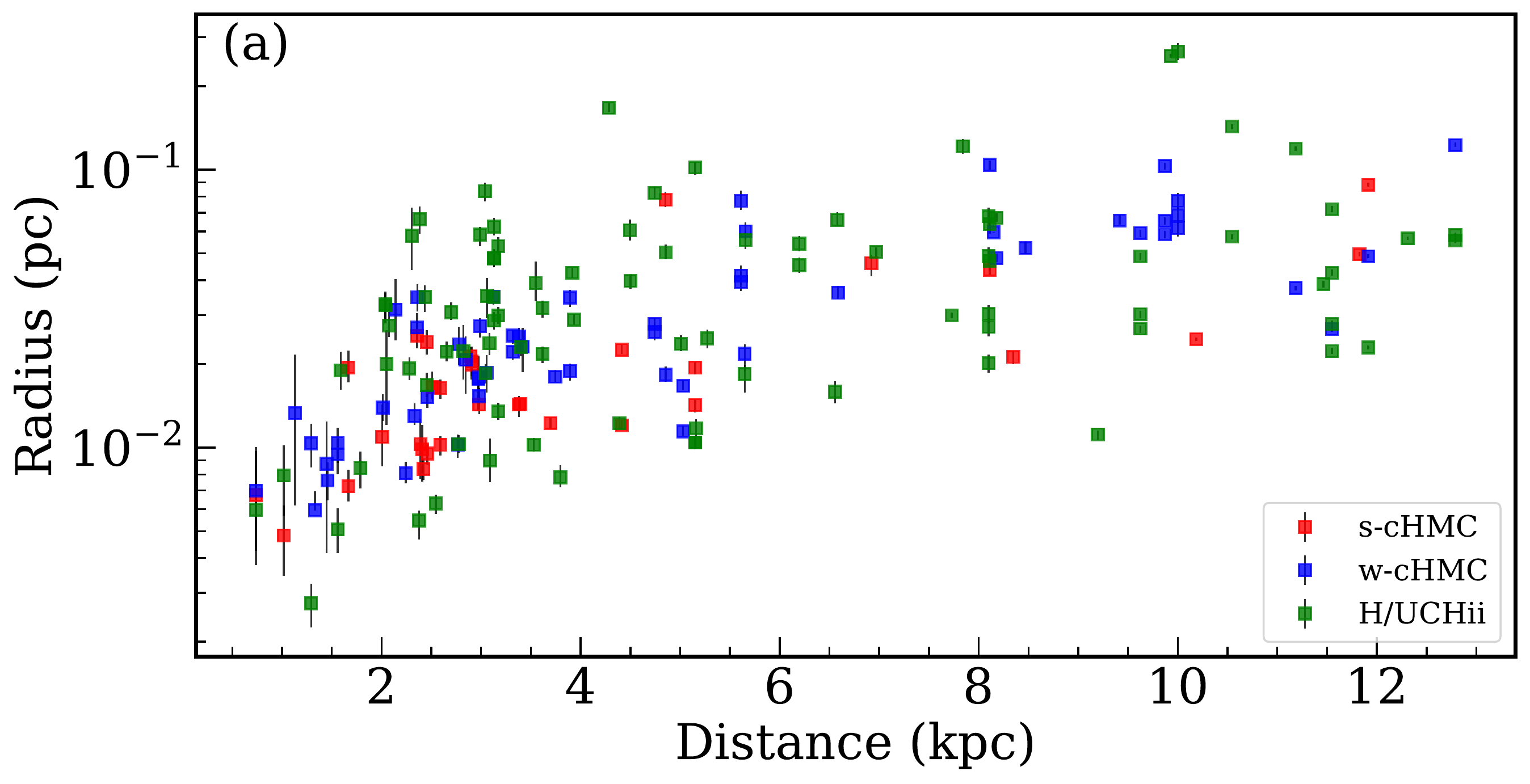}
\includegraphics[width=3.4 in]{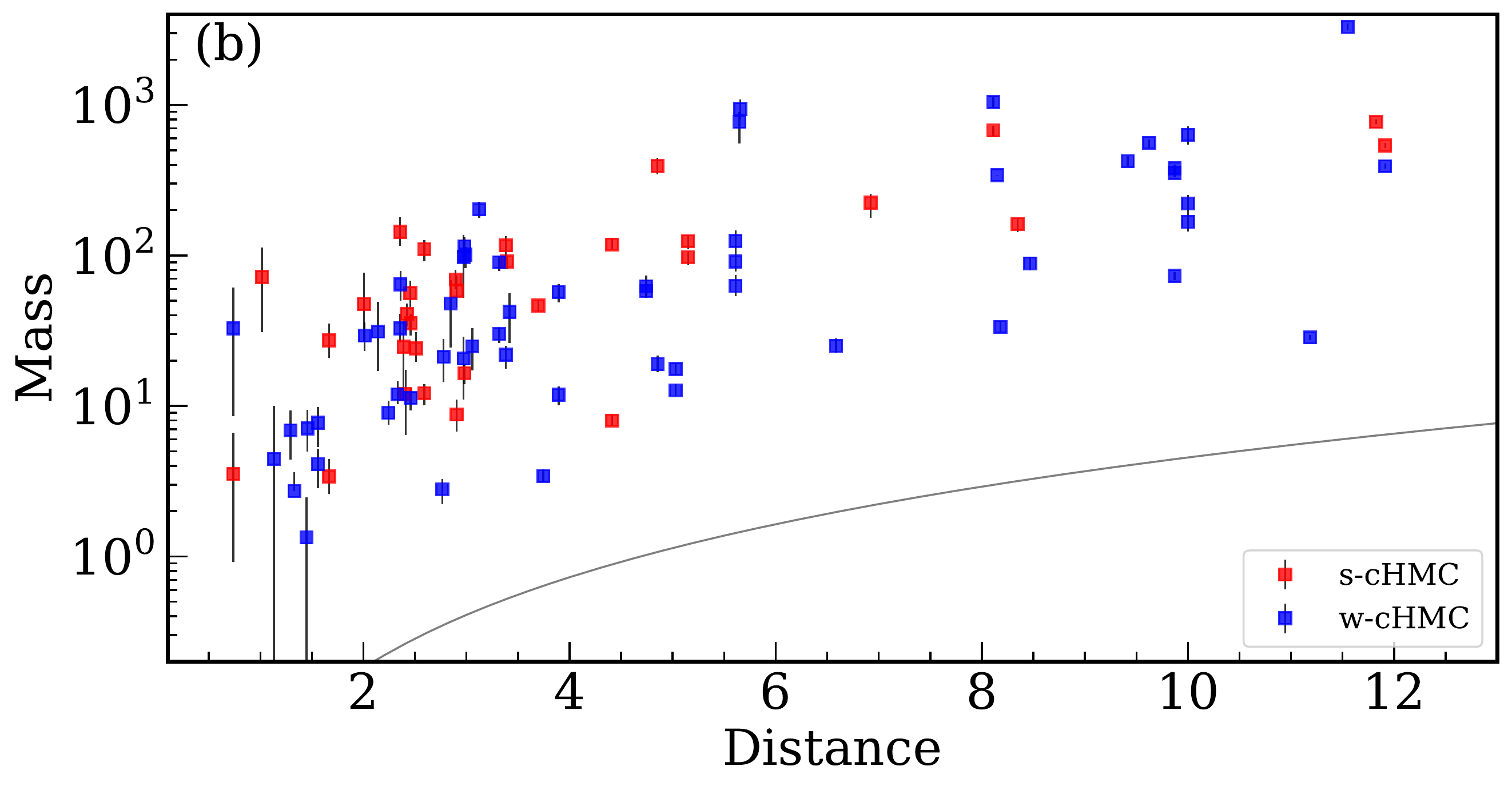}
\caption{(a):Radius versus distance for w-cHMC, s-cHMC, and \chii\ cores.  The p value of the KS test 0.03 suggests that the size distributions of w-cHMC and
s-cHMC come from an identical distribution. (b): Mass versus distance for w-cHMC, and s-cHMC cores. \chii\ cores are not shown
here since their masses cannot be estimated from 3\,mm continuum due to free-free contamination  (see text in Sect.\ref{sec:core:classify-HMC}). The curve is the mass sensitivity distribution as a function of the distance.
}
\label{fig:dist-dependence}
\end{figure*}

 Figure \ref{fig:dist-dependence} shows how the radii (R) and masses (M)  depend on distances (D) for the three groups (\PCHMC, \CHMC\ and H/UC-H{\sc ii}) of dense cores. There are clear increasing trends in R {\it vs.} D and M {\it vs.} D relations, further suggesting that distant sources were not well resolved. With future higher resolution observations, we expect that these distant cores with sizes of $\sim0.1$\,pc would break down to smaller sub-cores or condensations with sizes of $\sim0.01$\,pc. 
 Moreover, the masses of the \PCHMC\ and \CHMC\ cores can be found in Figure \ref{fig:dist-dependence}b far above the detection limit, which is in agreement
 with the nature of bright emission in continuum for those two types of cores.

Moreover, beam dilution impairs detection of hot and \chii\ cores at large distances.
 Therefore, we assume a smaller detection rate of such cores at larger distances. We tested this assumption by re-examining the detectablity of the COM and \halpha\ lines after pushing the observations of the relatively nearby sources to a large distance.
 In practice, we smoothed the cubes of SPWs7--8 for 10 cHMCs and 5 \chii\ cores at distances $\le2$\,kpc to the larger beam size that was transferred from the ATOMS typical angular resolution ($\sim 1.6$\arcsec) at a fixed far distance of 5\,kpc. In consequence, only 30\% of the selected cHMC sources were re-identified using the same HMC identification method as in Sect.\,3.2.2, while  80\% of the selected \chii\ sources have been re-identified. This result suggests that beam dilution affects the 
 detectabiliy for cHMCs more than it does for \chii\ cores at the ATOMS sensitivity.

\begin{figure*}
\centering
\includegraphics[width=3.4 in]{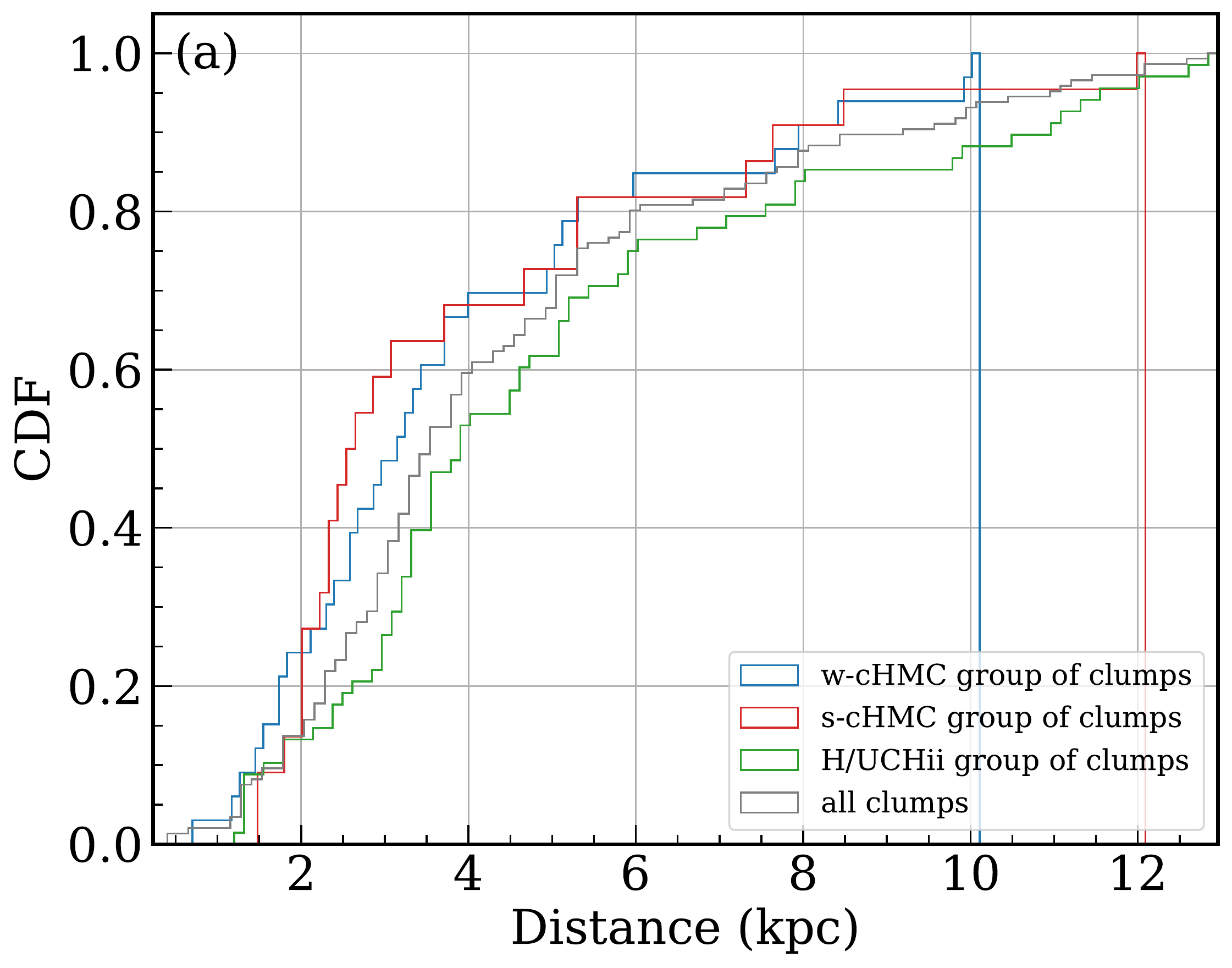}
\includegraphics[width=3.4 in]{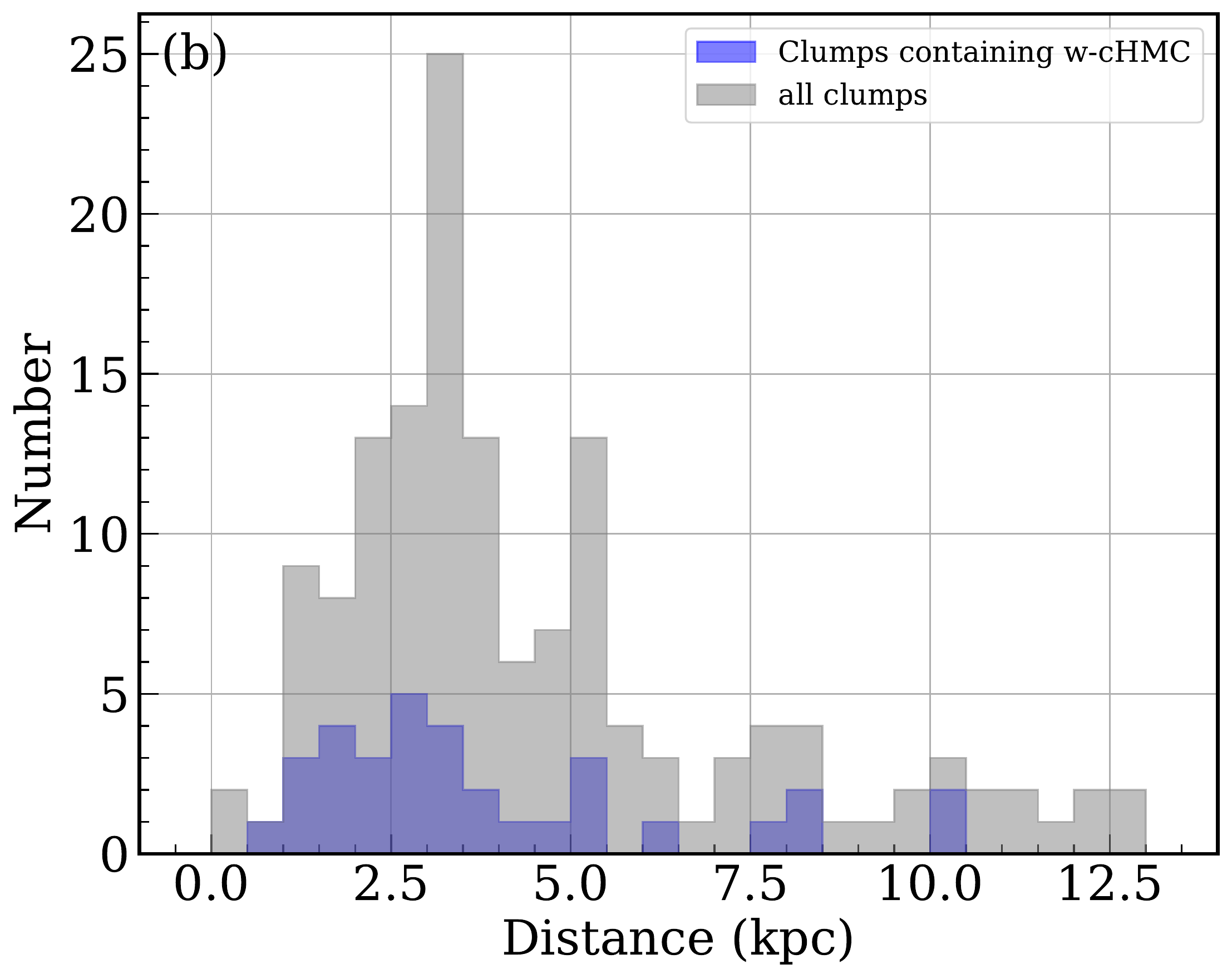}
\includegraphics[width=3.4 in]{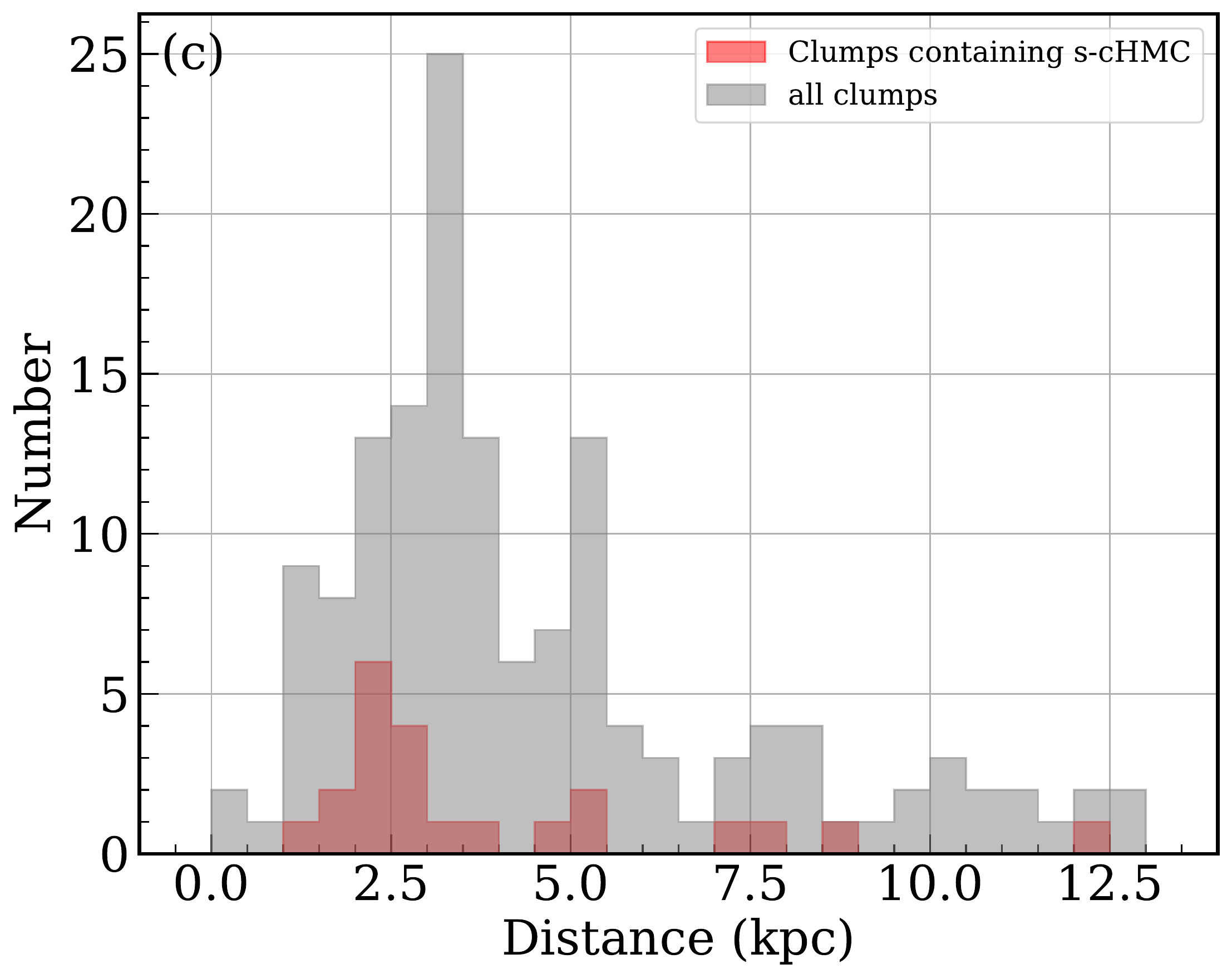}
\includegraphics[width=3.4 in]{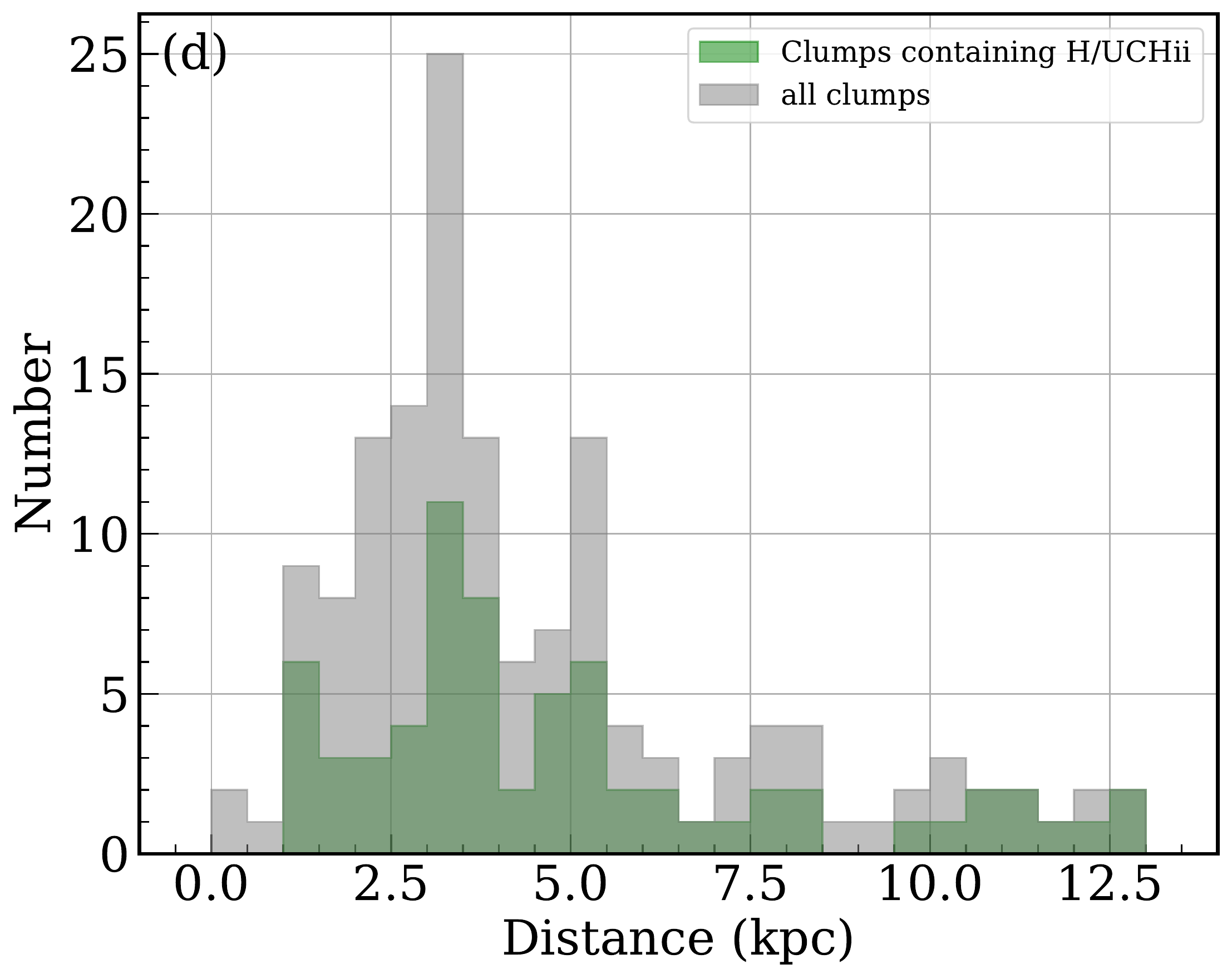}
\caption{ Distance distribution of the four groups of clumps. (a): cumulative distribution function. (b--d): histograms of all clumps against w-cHMC group,
s-cHMC group, and \chii\ group of clumps, respectively.
}
\label{fig:core-radius-kdeplot}
\end{figure*}

 Furthermore, Fig. \ref{fig:core-radius-kdeplot} shows the distance distributions of three groups of clumps, which contain \PCHMC, \CHMC\ and H/UC-H{\sc ii}. For comparison, we also plot the distance distribution of the full sample. Indeed, we find that the candidate hot cores  (\PCHMC\ and \CHMC) tend to be located at nearer distances (Fig.\,\ref{fig:core-radius-kdeplot}b,c), with their cumulative distribution curves (in red and blue colors in Fig.\,\ref{fig:core-radius-kdeplot}a) above the curve (in gray color) for the full sample.

We test whether the distance distributions of clumps containing different kinds of dense cores follow
the same distribution as the full sample with Kolmogorov–Sminov test. The unknown hypothesis is that two distributions follow the same
distribution. The tests are consistent with the unknown hypothesis  for the w-cHMCs group (P-value=0.37) and H/UC-H{\sc ii} group (P-value=0.67) due to their P-values being greater than 0.05, suggesting that the detection rates of w-cHMCs and H/UC-H{\sc ii} are not greatly affected by the distance issue. In contrast, the P-value for s-cHMCs is 0.06, indicating that the detection of s-cHMCs is significantly limited by distances. It may imply that the detection rate of s-cHMCs is significantly underestimated for distant sources. This is in line with  the cumulative distribution function of distances (Fig.\,\ref{fig:core-radius-kdeplot}a). In the figure,  clumps with s-cHMC appear to show smaller distances ($\sim2.6$\,kpc on average) than those ($\sim 3.2$\,kpc on average) with w-cHMC, indicating that the detection of hot cores at large distances is incomplete.

\subsection{Nature of the COM-containing  and not \chii-associated cores}
\label{sec:disc_COMnature}
\begin{figure}
\centering
\includegraphics[width=3.4 in]{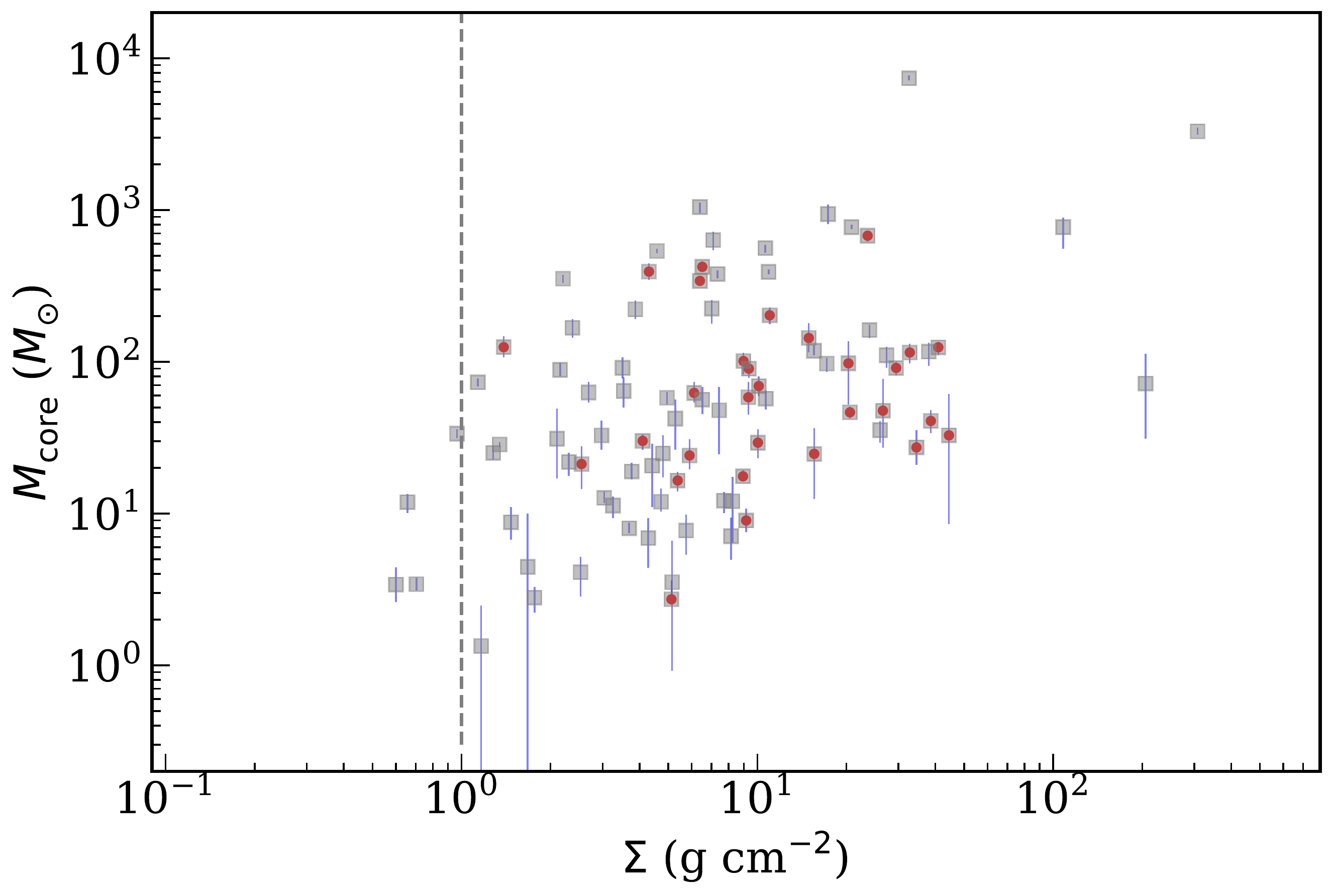}
\caption{Distributions of the mass and mass surface density of the 90 COM-containing cores that are not associated the \chii\ signature.
The two parameters were calculated assuming a temperature of 100\,K typical of HMCs (see Appendix\,\ref{sec:app-mass}).
The dashed vertical line indicates a stringent
theoretical threshold of 1\,g~\cmcm\ above which the cores most likely form
high-mass stars.  The red dots highlight the cores associated with the Class\,II \chiiioh\ masers, showing  that they are widely distributed, but  above the 1\,g~\cmcm\ line. 
}
\label{fig:core:paras}
\end{figure}

Since emission in the COM transitions covered by the ATOMS observations can be detected towards 
several distinct objects, including low/high-mass prestellar cores, HMPO cores, HMCs, and even hot corinos associated with 
low-mass protostars \egcite{Vas14,Jim16,Som18,Cse19,Mol19,Jor20,Hsu20}, it is necessary to constrain the nature of the 90 COM-containing cores that are not associated with \chii\ (see Sect.\,\ref{sec:core:classify-HMC}). 
To this end, we investigate the distribution of both the mass ($M_{\rm core}$) and the mass surface density 
($\Sigma_{\rm core}$) of the 90 COM-containing cores, as shown in Fig.\,\ref{fig:core:paras}.
They have a  median value of mass equal to $\sim57$\,\msun\ in an 
IQR
of 19--138\,\msun, and surface densities with a median value of $\sim6.5$\,g~\cmcm\ in an IQR of 3.3 to 15.6\,g~\cmcm.

Theoretically, cores (on scale $<0.1$\,pc) with $\Sigma_{\rm crit}\ge1$\,g~\cmcm\  will most likely form high-mass stars \citep{Kru08}.
Searching from existing maser catalogue\footnote{\url{http://maserdb.net/download.pl}}\citep{Lad19},  we find that $\sim 34\%$ of the 90  COM-containing  cores are associated with Class\,II \chiiioh\ masers located within the core size (red dots in Fig.\,\ref{fig:core:paras}).  Class\,II \chiiioh\ masers have been found to be exclusively associated with high mass star forming regions \egcite{Cyg09}. The spread of $\Sigma_{\rm core}$
for the cores associated with the \chiiioh\ masers is consistent with the theoretical prediction (i.e., the red dots versus 
the vertical dashed line in the figure).
Following the threshold $\Sigma_{\rm crit}\ge1$\,g~\cmcm,
we find that $\sim 96\%$ of the 90  COM-containing cores above it 
will most likely form high-mass stars. We note that there are six  COM-containing cores, three of which are  associated with Class\,II masers, above the threshold but  
with $M_{\rm core}<10$\,\msun. We suggest that these cores could have been in the process of forming high-mass stars for a significant fraction of their lifetime, so that a large fraction of their mass has been accreted onto the central HMPOs.
We also note that if  the actual temperature of the 90 COM-containing cores is lower than 100\,K, the possibility that they form high-mass stars will be higher. For reference, if the temperature is assumed to be 50\,K, the mass and surface density will be about
two times higher. 

Overall, the probability that the identified COM-containing cores are associated with  low-mass star forming objects (i.e., low-mass prestellar cores, and hot corinos) is rather low. Given that emission in the COMs transitions available in the ATOMS data has been already observed towards other HMPO and HMC regions \citep{Cse19,Jor20}, we conclude that most (if not all) of the 90  COM-containing cores should be either HMPO cores or HMCs. 
Note that HMPOs are simply assumed here to be 
younger than HMCs although this timescale relation has yet to be determined.

To distinguish definitely between  the two types of high-mass star-forming cores requires knowledge of their temperatures. The temperature  is usually computed from a
rotational diagram analysis of multiple transitions of one or more COM species (e.g., \chthocho, and \chiiioh). However, the range 
of upper level energy of the COM lines covered in the ATOMS survey is too narrow to permit a reliable temperature determination. This thus requires follow-up observations of higher frequency transitions that arise from higher energy levels.  For simplicity, hereafter, we  treat the 90 COM-containing cores as cHMCs (see Sect.\ref{sec:core:classify-HMC}) since the detection rate of COMs towards HMCs is presumably higher than towards HMPOs.

In short, most of the 90 COM-containing cores are presumably in the very early stages of high-mass star formation. We include them in two separate catalogues: one including 32 \CHMC s and the other including 58 \PCHMC s, with the former catalogue presumably containing more robust HMCs candidates than the latter.
These  catalogues constitute a unique database of objects for  future follow-up higher-resolution observations to investigate the  dynamical processes related to high-mass star formation, such as infall, outflow, and rotation.

\subsection{Constraints on the Duration of the HMC Phase}
Because heating by the forming star precedes ionization, the
general picture is that HMCs precede \chii\ in the evolution of
a core to a massive star \citep{Gar19,Kur00,van04,Ces05,Rat11}. 
Our large sample, unbiased towards either HMCs or \chii\ cores, enables further study. First,
the two phenomena are not exclusive, as there is a substantial
overlap between HMCs and \chii\ 
(Fig.\,\ref{fig:core-venn1}). 
Nearly half (48/89) of \chii\ in the sample show COMs, indicating that hot cores may persist during nearly half of \chii\ lifetime even when the central high-mass protostars start to ionize the surroundings.

Second,
with certain assumptions, we can use the numbers of candidate HMCs and cores with \chii\ to estimate the relative duration of the HMC phase. For a rigorous result, the survey must be complete, the
evolution through the phases must be a continuous process proceeding at a steady rate for a period longer than the last
phase considered, and other variables, such as the mass of the
star, cannot be important. Further study of this sample is 
needed to check these assumptions, so we offer here only some preliminary considerations.

 As discussed in Sect.\,\ref{sec:limitations}, the ATOMS survey is more complete in the detection of H/UC-H{\sc ii} regions than HMCs, so we have a lower limit on the number counts of HMCs.  
The total number (138) of COM-containing dense cores (w-CHMC and s-CHMC) is already larger than the number (89) of \chii\ cores. The number of strong candidate HMCs (56) is smaller than the number (89) of \chii\ cores but may be greatly underestimated due to the distance effect. The most important unknown is the mass of the forming star because the number of ionizing photons is a very strong function of stellar mass. If all these cores will form stars capable of producing
a \chii, the current lack of one is due to evolution.
Further study of the luminosity of the individual cores with JWST observations could test this assumption.

With all these caveats, we suggest that the duration of candidate hot cores  is at least comparable to the lifetime of \chii\ and is probably greater.

\subsection{The origin of chemical differentiation among high-mass star forming cores}

\begin{figure*}
\centering
\includegraphics[width=3.4 in]{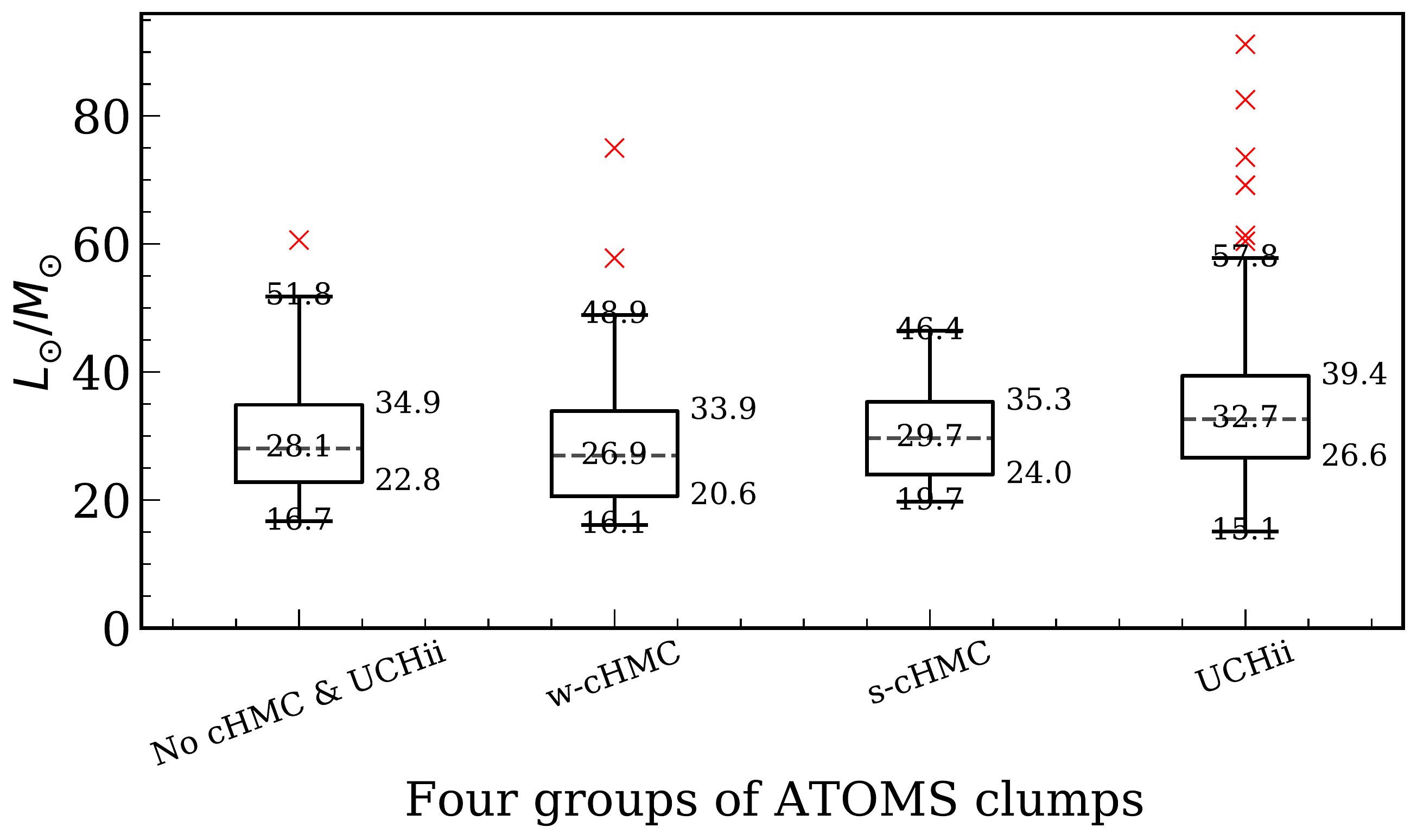}
\includegraphics[width=3.4 in]{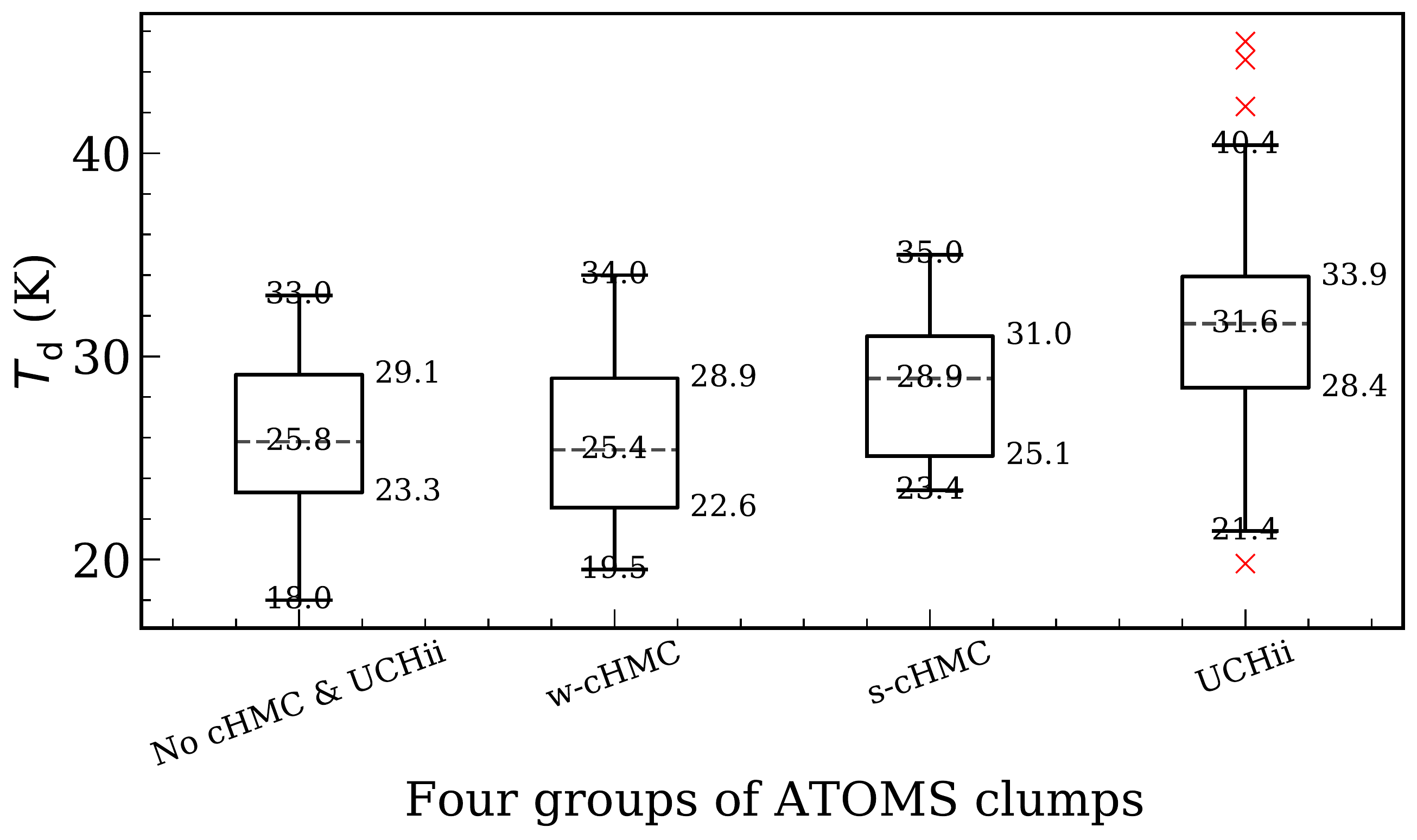}
\caption{Box and whisker plot showing for four groups of clumps the distribution of the L/M ratio (panel\,a),
 dust temperature $T_{\rm d}$ (panel\,b). The numbers associated with the boxes from the top to bottom represent the upper quartile, 
median (inside the box), and lower quartile, respectively. The red crosses indicate the outliers outside 1.5 times the interquartile range either 
above the upper quartile or below the lower quartile.
}
\label{fig:clump:statistics}
\end{figure*}

\begin{figure*}
\centering
\includegraphics[width=2.3 in, height=1.8 in]{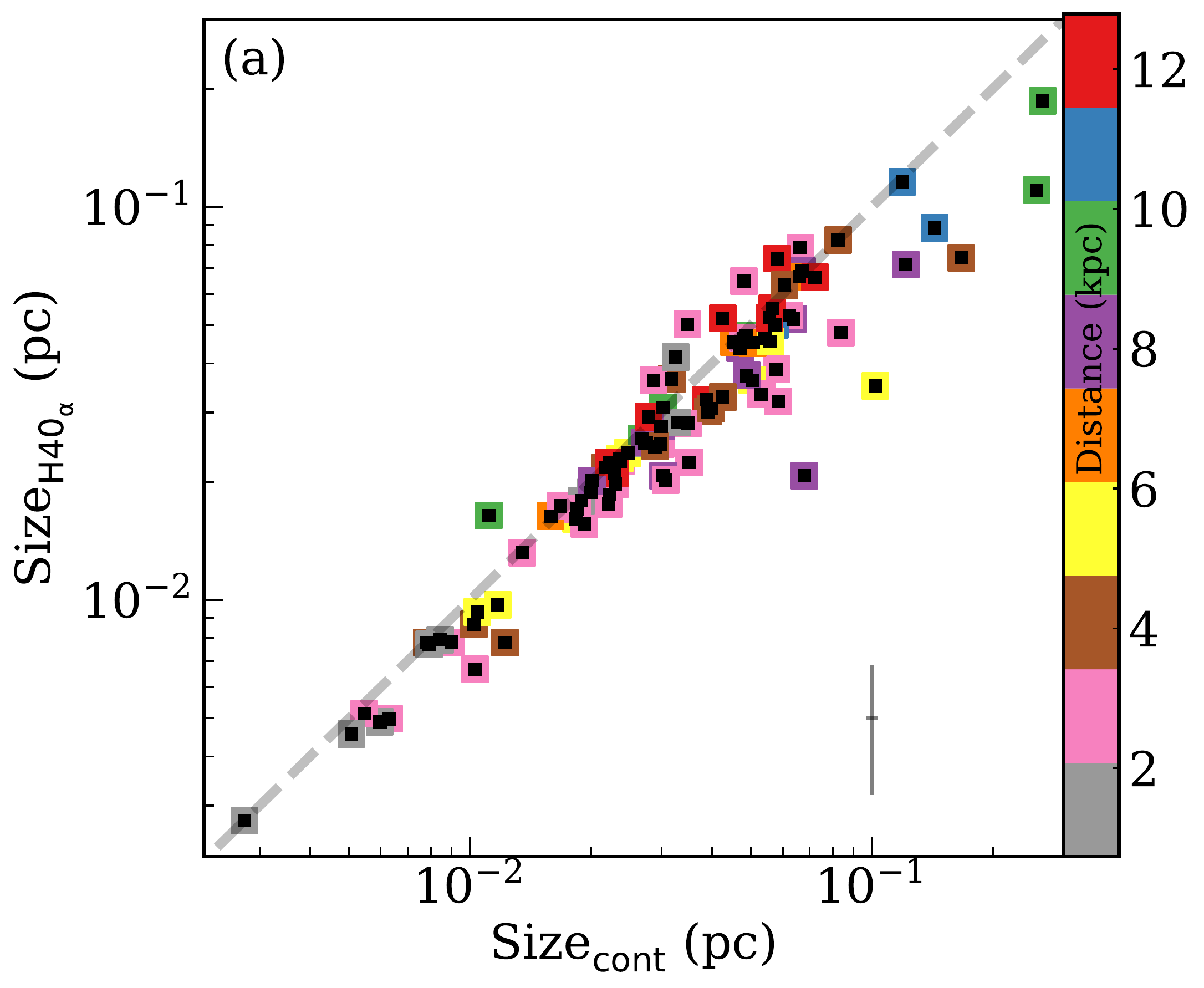}
\includegraphics[width=2.3 in, height=1.8 in]{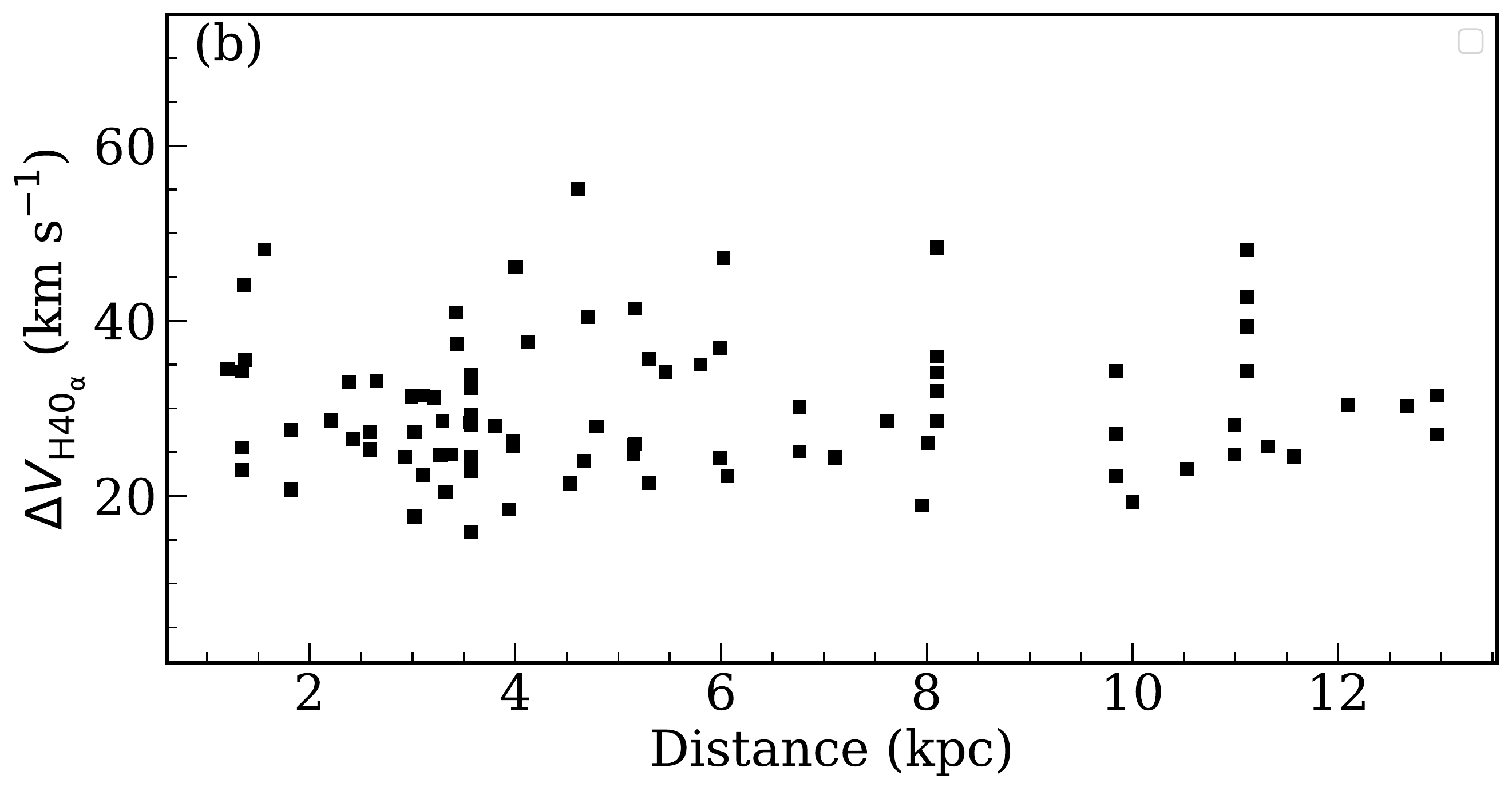}
\includegraphics[width=2.3 in, height=1.8 in]{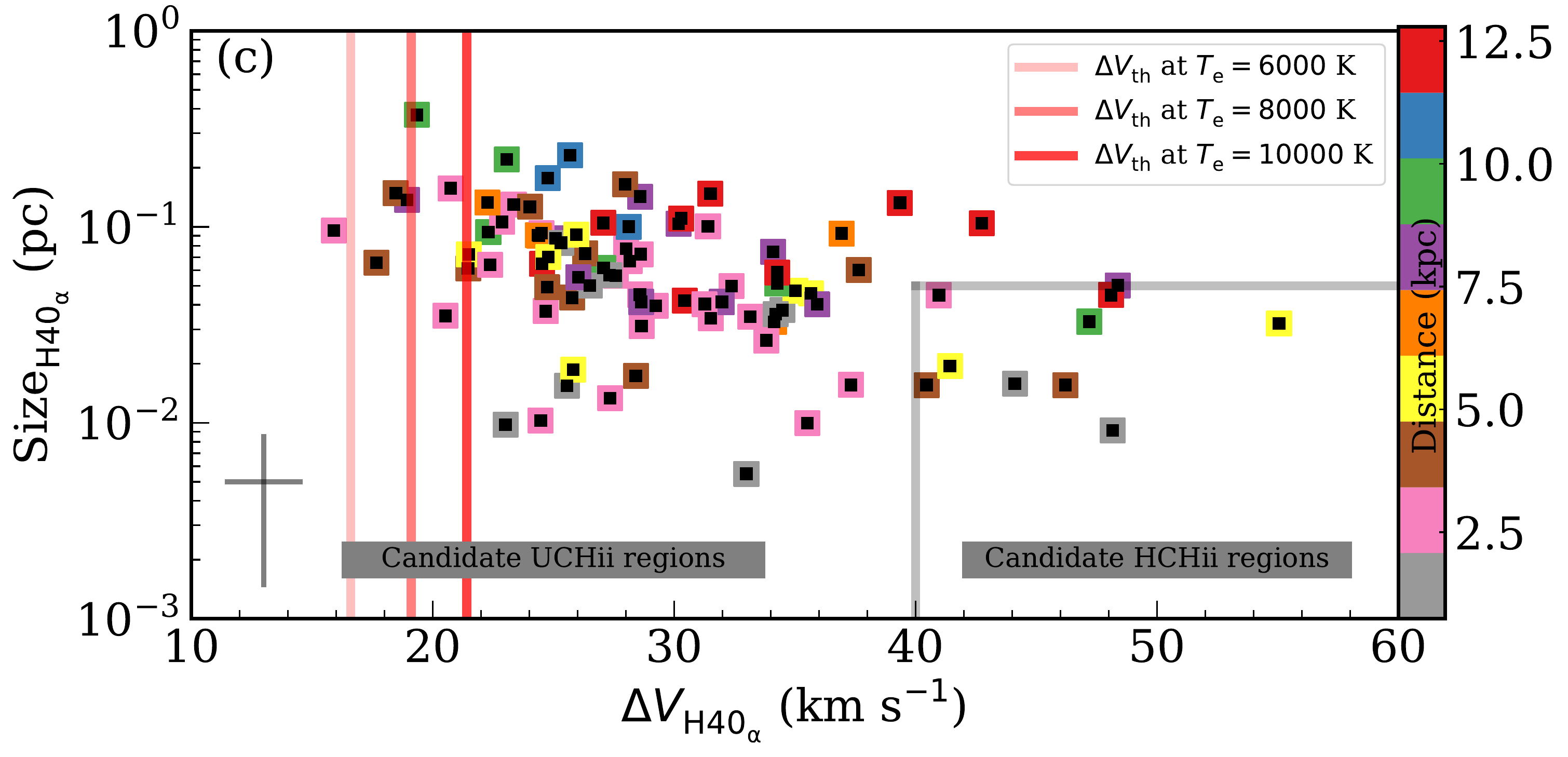}
\caption{
Size and \halpha\ line width of the 89 \chii\ cores. (a): comparison between the two types of core sizes, one measured from 3\,mm continuum (in x-axis) and
the other from \halpha\ emission (in y-axis). The color scheme encodes distance information of the cores. The typical error bars for the two
measurements are presented at the bottom right.
(b): distribution of the \halpha\ line width as a function of distance. (c):
Relation between the size of the \chii\ cores measured from \halpha\ emission and the line width of \halpha\ ($\Delta V_{\rm H40_{\alpha}}$). 
The color scheme is the same as in panel\,(a).
The \chii\ cores are roughly divided into 79 \uchii\ and 10 \hchii\ classes.
For reference, the thermal-broadening line widths are given in three color-gradient vertical lines at 16.6\,\vel, 19.1\,\vel,
21.4\,\vel\  for the electronic temperatures 6,000\,K, 8,000\,K, and 10,000\,K, respectively. The typical error bars for the two parameters are presented at the bottom left.
}
\label{fig:uhcii-linewidth-radius}
\end{figure*}

Some high-mass star forming cores are found to be chemically rich while others are not. Even cHMCs show differing richness in molecular lines especially in COMs. The median $N_{\rm line}$ values are 8 within an IQR of [5, 16] for the 58 \PCHMC s, and 57 within 
an IQR of [28, 123] for the 32 \CHMC s. Would this different chemical richness (i.e., $N_{\rm line}$) 
be related to an evolutionary sequence from \PCHMC s$\rightarrow$~\CHMC s?

The chemical differentiation among high-mass star forming cores could be partially attributed to an evolutionary effect. More evolved cores
with higher luminosity can  heat a larger volume of their surroundings, evaporating molecules from ices and thus increasing the gaseous molecular abundances.

The 146 analyzed ATOMS clumps can be classified into four main groups 
according to the types of cores they harbour: (i) the \PCHMC\ group with 28 clumps hosting \PCHMC\ cores,
(ii) the \CHMC\ group with 17 clumps hosting \CHMC\ cores, (iii) the \chii\ group with 
68 clumps containing \chii\ cores, and (iv) the ``unknown'' group with 28 clumps without 
cHMC or \chii\ cores. There are five clumps with both \PCHMC\ and \CHMC\ cores. 
Note that if a clump hosts both \chii\ core and cHMC, it will be placed into the \chii\ group. 
In addition, the \PCHMC\ group
could contain \CHMC\ cores that are not detectable yet or appear as \PCHMC\ cores due to the limited angular resolution and sensitivity. That is, there could be an underlying overlap between \PCHMC\ and \CHMC\ groups of clumps, so the following analysis should be treated with caution.
To explore the evolutionary stage of the clumps in the different groups we resort to two parameters available 
for the ATOMS clumps (see Paper\,I):  the bolometric luminosity to mass ratio $L_{\rm bol}$/$M_{\rm clump}$ and the 
dust temperature $T_{\rm d}$. These parameters, which cannot be directly determined for the cores with the ATOMS data alone, 
are distance independent and have been proposed as indicator of the evolutionary stage of clumps \egcite{Guz15}.

Figure\,\ref{fig:clump:statistics} plots the distribution of the two parameters for the four groups of 
clumps. The ``unknown'', \PCHMC, \CHMC, and \chii\ groups have 
respective median $L/M$ values 28.1, 26.9, 29.7, 32.7\,\lsun/\msun. These values fall into 
the range of $2\le L/M\le40$\,\lsun/\msun\ observationally defined for the high-mass star-forming objects \citep{Gia17,Eli17}. 
This result suggests that the four groups of clumps harbour embedded phases of high-mass star formation. As such, many of
the clumps in  the ``unknown'' group could not be really devoid of high-mass star-forming cores, and instead harbour HMPO objects or even more evolved HMC  that could not be detected with the sensitivity of the ATOMS data. If we only consider the \PCHMC, \CHMC, and \chii\ groups, 
there is a weak increasing trend in the median values of the $L_{\rm bol}$/$M_{\rm clump}$ ratio.
A similar, but stronger, trend also appears in the temperature distribution among these three groups of clumps.

The differences in $L_{\rm bol}$/$M_{\rm clump}$ and $T_{\rm d}$ of the natal clumps are not very significant. If the evolution
from HMCs to \chii\ is correct for individual cores, the explanation for the poor correlation with clump-scale properties may
be that the evolution of individual cores to stars within a clump differs, and the clump-scale evolution of $L_{\rm bol}$ is 
driven by the most massive/evolved core. To study evolution at the core level, higher angular resolution and sensitivity in the mid-infrared to far-infrared is needed to pinpoint the embedded protostars, which cannot be well resolved at the current resolution of the infrared data, through the spectral energy distribution over a large dynamical frequency range.

In addition, recent studies of low-mass cores indicate that chemical richness of star forming cores may not be related to evolutionary sequence. \citet{Bel20}  suggested a luminosity threshold of 4\,$L_{\odot}$, above which low-mass cores should exhibit spectral features from at least one COM and may contain hot corinos. However, in a recent study of a large sample of 49 low-mass star forming cores located in the Orion complex by \citet{Hsu20}, many protostellar objects with luminosities much greater than 4\,$L_{\odot}$ do not show any clear sign of COMs or hot corinos. This indicates that the chemical differentiation among star forming cores depends on their  chemical history, as well as their luminosity. 

An alternative scenario could be that some star forming cores in their ``cold" core phase form abundant molecules such as COMs on ice mantles through grain-surface chemistry while other cores do not. These different behaviours are probably related to varying physical and chemical properties as well as initial conditions of different sources.
When molecules evaporate from ice in the ``hot" core phase, their gaseous molecular abundances should also vary greatly.  This scenario could be tested by detailed chemical studies toward a sample of ``cold clumps" like in infrared dark clouds that lack active star formation and may represent the very initial conditions of high-mass star formation. 
Detailed chemical investigations are beyond the scope of this work. More detailed chemical studies of molecular abundances of dense cores will be presented in forthcoming papers.

\subsection{Size versus \halpha\ linewidth relationship for \chii\ cores}
\label{sec:disc:size-FWHM-ha}

The 89 \chii\ cores could be associated with either \hchii\ or \uchii\ regions, and we  attempt to distinguish them by their 
observable properties (see Sect.\,\ref{sec:intro}).
To this end, we  focus on the size and line width of the radio recombination line ($\Delta V_{\rm H40_{\alpha}}$) of the \chii\ cores,
as shown in Fig.\,\ref{fig:uhcii-linewidth-radius}.
These cores have two types of sizes, one measured from 3\,mm continuum and the other one from \halpha\ emission. As shown in Fig.\,\ref{fig:uhcii-linewidth-radius}a, the two types of sizes match each other well until the size is greater than 0.1\,pc. This result suggests that
\halpha\ emission mainly contributes to 3\,mm continuum when the \chii\ core size is less than 0.1\,pc. From distance information of the cores as indicated
in the colorscale in Fig.\,\ref{fig:uhcii-linewidth-radius}a, it can be seen that the deviation of the two types of core sizes does not depend on the distance. We suspect that when the \chii\ core size is greater than 0.1\,pc, more cold dust emission could be included within it, leading to an overestimation of the size of the real ionized gas region. In view of this, the size measured from \halpha\ emission for the 89 \chii\ cores will be discussed in the following. These cores  have sizes from $\sim 5.3\times10^{-3}$ to $3.7\times10^{-1}$\,pc with a median value of $5.7\times10^{-2}$\,pc. 

The  $\Delta V_{\rm H40_{\alpha}}$ of the 89 \chii\ cores range from $\sim 15$ to 55\,\vel\ with 
a median value of 28\,\vel. 
The measurement of $\Delta V_{\rm H40_{\alpha}}$ is not found to be distance-dependent, as shown in  Fig.\,\ref{fig:uhcii-linewidth-radius}b.
Actually, the $\Delta V_{\rm H40_{\alpha}}$ values are typical of both \uchii\ and \hchii\
regions, namely in a range of $\sim10$ to 40\,\vel\ for the former and 40 to 60\,\vel\ or even higher for
the latter \egcite{Hoa07,Riv20}. Following the typical size and line width of radio recombination lines of \chii\ regions (see above), 
the 89 \chii\ cores can be divided into 79 \uchii\  and 10 \hchii\ cores (see Fig.\,\ref{fig:uhcii-linewidth-radius}c). 
That is, most ($\sim 90\%$) of the compact-\Hii\ cores are \uchii\  cores while only a small fraction ($\sim 10\%$) of them are \hchii\ cores. Therefore, if the lifetime of  \uchii\ is of the order of $10^{5}$\,yr, the lifetime of \hchii\ will be $\sim 10^{4}$\,yr \egcite{Woo89, Dav11}.

In general, the widths of radio recombination lines are due to a combination of thermal, dynamical, and/or pressure broadening mechanisms \egcite{Ket08,Ngu17, Riv20}. The thermal broadening, $\Delta V_{\rm th}$,  is given  by  $\Delta V_{\rm th}=({\rm 8ln2}k_{\rm B}T_{\rm e}/m_{\rm H})^{1/2}$ where $k_{\rm B}$ is the Boltzmann constant and $m_{\rm H}$ is the mass of the atom hydrogen.
For electron temperatures of $T_{\rm e}$ = 6,000\,K, 8,000\,K, and 10,000\,K, the respective thermal widths are 16.6\,\vel, 19.1\,\vel,
21.4\,\vel, (see the three vertical color-gradient lines in Fig.\,\ref{fig:uhcii-linewidth-radius}c). Most ($\sim 96\%$) of the \chii\ cores are found to have $\Delta V_{\rm H40_{\alpha}}$ widths larger than $\sim 19$\,\vel, which corresponds to the thermal broadening for an electron  temperature of 8,000\,K typical of an ionized plasma \egcite{Ost06}. This result indicates that in addition to the thermal broadening, non-thermal (e.g., dynamical and/or pressure) broadening mechanisms are responsible for the observed large $\Delta V_{\rm H40_{\alpha}}$ in most of the 89 \chii\ cores. 

Moreover, we find that the \halpha\ line width decreases as the size of the core increases. To investigate if the trend is distance-dependent or not, we present the distance information of each \chii\ core in colorscale in Fig.\,\ref{fig:uhcii-linewidth-radius}. The figure shows that the decreasing trend persists rather than breaks at
different distances, and thus it is physically meaningful. It indicates that non-thermal (e.g., dynamical and/or pressure)  mechanisms dominate the broadening on smaller scales of 
the \chii\ cores \citep{Gar19}.


\section{Summary and conclusion}

Using the 3\,mm  ATOMS survey data, we searched for the presence of compact cores towards 146 IRAS clumps  located within the inner galactic disk  in the range $-80\degr<l<40\degr$ and $|b|<2\degr$. The compact cores were extracted from 3\,mm 
continuum emission maps through a combination of the {\it Dendrogram} algorithm and CASA-{\it imfit} task. The main results of this work are summarized as follows:
\begin{itemize}

     \item We compiled three catalogues of compact cores with different characteristics. One catalogue, referred to as the \chii\ catalogue, includes 89 cores that enshroud \chii\ regions as indicated by the presence of compact \halpha\ emission; a second catalogue, refereed as the pure \CHMC\ catalogue, includes 32 \CHMC s associated with a rich spectrum ($N\ge20$ lines)  of  complex organic molecules (COMs); the third catalogue, referred to as the pure \PCHMC\ catalogue includes 58 \PCHMC s with relatively low levels of COM richness (i.e., $5\le N_{\rm line}<20$). These three catalogues of compact cores provide a crucial foundation for future studies of the early stages of high mass star formation across the Milky Way. We find that nearly half of \chii\ cores are candidate hot molecular cores. 

     \item For completeness, another catalogue that includes the remaining 274 cores
	without the signature of either cHMC or \chii, refereed as the ``unknown'' catalogue, is also provided. These cores  deserve future detailed investigation, since some of them could be in the 
	evolutionary stages of high-mass star formation prior to formation of a HMC, and thus could be good targets for studying initial conditions of high-mass star formation.
  
 \item  The detection rate of candidate distant hot cores is greatly underestimated due to relatively poorer linear resolution and line sensitivity than for nearer sources. However, the total number (138) of the detected COM-containing dense cores is still larger than the number (89) of \chii\ cores, indicating that the duration of high-mass protostellar cores (or cores) showing line-rich features could be at least comparable to the lifetime of \chii\ phase. 
     
 \item  We find weak increasing trends in $L/M$ ratio and dust temperature (T$_d$) for the natal clumps that contain different kinds of dense cores (\PCHMC$\rightarrow$\CHMC$\rightarrow$\chii). However, 
 the chemical classes indicating core evolution are not well traced by the $L_{\rm bol}$/$M_{\rm clump}$, suggesting that the evolution of individual cores is not captured by that evolutionary indicator used for clumps.  
     
     \item Based on the relationship between size and \halpha\ line width, we conclude that of the 89 \chii\ cores,  79 are associated with \uchii\ and 10 with  \hchii\ regions.  We also find that the width of the \halpha\ line increases as the core size decreases, suggesting that non-thermal (e.g., dynamical and/or pressure) line-broadening mechanisms are dominant on the smaller scales of the H/UC-HII cores.

\end{itemize}

The 3\,mm interferometric catalogues of high-mass star-forming cores provided here can be used to study the kinematics and dynamics of molecular gas in the vicinity of newly-formed (proto)OB stars and somewhat-evolved \chii\ regions.
Such studies will definitely advance our understanding
of the dynamical processes in the early stages of high-mass star formation, such as infall, outflow, and rotational motions, and our knowledge of the roles of feedback (e.g., outflows, stellar winds, and \chii\ regions) in the process of high-mass star-formation.

\medskip
\noindent{\textbf{Acknowledgements}}\\
We thank the anonymous referee for comments and suggestions that greatly improved the quality of this paper.
H.L. Liu acknowledges funding from a Fondecyt Postdoctoral project (code 3190161).
H.L. Liu also thanks Shanghai Astronomical Observatory for providing office facilities that facilitated analysis of this work.
T. Liu acknowledges the supports from the international partnership program of Chinese academy of sciences through grant No.114231KYSB20200009, 
National Natural Science Foundation of China (NSFC) through grant NSFC No.12073061, and Shanghai Pujiang Program 20PJ1415500.
G.G. acknowledges support from ANID project AFB 170002. S.-L. Qin is supported by NSFC under No.12033005.
L.B. acknowledges support from CONICYT project Basal AFB-170002.
AS gratefully
acknowledges funding support through Fondecyt Regular (project code
1180350) and from Chilean
Centro de Excelencia en Astrofísica y Tecnologías Afines (CATA) BASAL
grant AFB-170002. RA acknowledges funding support from CONICYT Programa  de  Astronom\'ia  Fondo  ALMA-CONICYT  201731170002.
This work was carried out in part at the Jet Propulsion Laboratory, which is operated for NASA by the California Institute of Technology.
CWL is supported by the Basic
Science Research Program through the National Research Foundation
of Korea (NRF) funded by the Ministry of Education, Science and Technology (NRF-2019R1A2C1010851).
K.T. was supported by JSPS KAKENHI Grant Number 20H05645.
This paper makes use of the following ALMA data: ADS/JAO.ALMA\#2019.1.00685.S. ALMA is a partnership of ESO (representing its member states), NSF (USA), 
and NINS (Japan), together with NRC (Canada), MOST and ASIAA (Taiwan), and KASI (Republic of Korea), in cooperation with the Republic of Chile. The Joint 
ALMA Observatory is operated by ESO, AUI/NRAO, and NAOJ.
This research made use of astrodendro, a Python package to compute dendrograms of Astronomical data ({\url{http://www.dendrograms.org/}}).
This research made use of Astropy,
a community-developed core Python package for Astronomy (Astropy
Collaboration, 2018).

\noindent{\textbf{Data availability}}\\
The data underlying this article are available in the article and in its online supplementary material.

\vspace{-5mm}


\input ourwork.bbl
%
\appendix
\section{Calculation of distances}
\label{sec:app-kin-dist} 

\begin{figure}
\centering
\includegraphics[width=3.4 in]{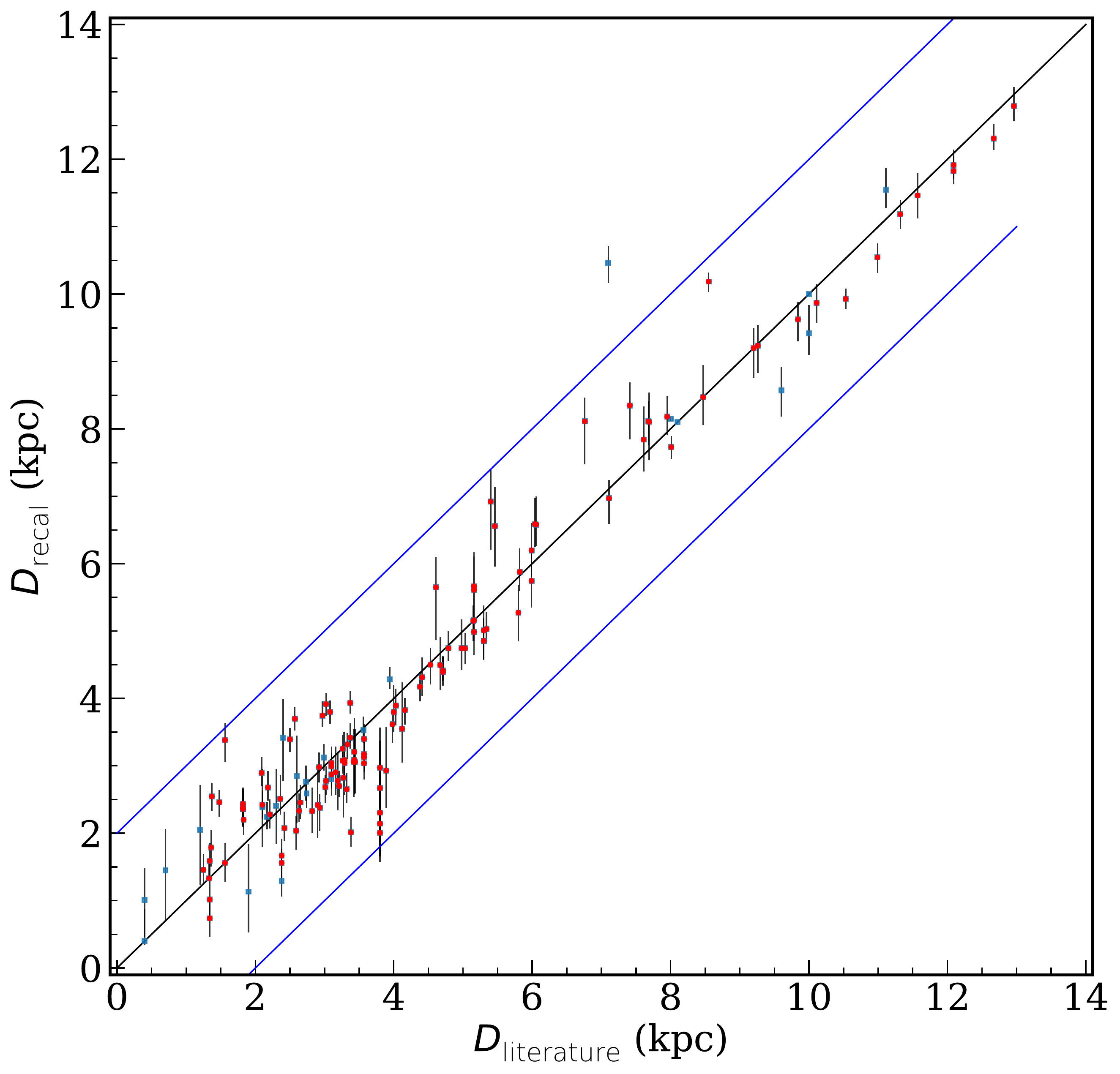}
\caption{Comparison between the distances calculated in this work and those from the literature for the 146 ATOMS clumps. The black and blue lines indicate unity and 2.0\,kpc offsets, respectively. 
}
\label{fig:app:dist}
\end{figure}

Due to the lack of distance uncertainty information from the literature \citep{Fau04, Urq18} for the 146 target clumps, their kinematic distances were recalculated  using the Monte Carlo kinematic distance tool developed by \citet{Wen18}. 
With the Galactic coordinates and radial velocity parameters of a source inputted, the tool can return the distances and the uncertainties for the source leveraging the rotation curve of \citet{Rei14}  with the updated solar motion parameters \citep{Rei19} and the Monte Carlo technique with 10,000 samplings. The radial velocity measurements by \citet{Liu16} rather than by
\citet{Bro96} were adopted in this work since the velocity resolution (0.42\,\vel) in \citet{Liu16} at which the measurements were made  better than that (0.76\,\vel) in \citet{Bro96}.

Apart from the two sources located near the Galactic center, kinematic distance solutions have been calculated for 144 ATOMS clumps. The distance ambiguities (if any) have been resolved following \citet{Fau04, Urq18}. Moreover, the kinematic distance solutions have been improved for the three sources using maser parallax distances available in the literature \citep{Rei14,Urq18}. Overall, the typical relative distance uncertainty for the 144 ATOMS clumps is $\sim 7\%$. For the remaining two sources located near the Galactic center, their kinematic distances could not be calculated from the rotation curve and thus the values from the literature along with a typical relative distance uncertainty of $\sim 7\%$ have been adopted. The recalculated distances are listed in Table\,\ref{tbl:app:dist}.

Comparing the recalculated distances with those from \citet{Fau04, Urq18}, we find a good agreement (Pearson correlation coefficient  $\rho$ is 0.97) as shown in Fig.\,\ref{fig:app:dist}. In addition, the relative distance uncertainties are found to have an IQR of $\sim$~[$5\%$, $12\%$] with a median value of $\sim 7\%$.

\begin{table*}
\label{tbl:app:dist}
	\centering
	\caption{Distances of 146 ATOMS clumps.}
	\label{atomsample}
	\resizebox{18cm}{!}{
	\begin{tabular}{cccccccccccc} 
		\hline

 ID &             Name &              $l$  &             $b$ &   $V_{\rm lsr}$  &             $D$  &  ID &             Name &              $l$  &             $b$ &   $V_{\rm lsr}$  &             $D$  \\
    &                  &                 deg  &    deg & km s$^{-1}$  &    kpc   &     &                  &                 deg  &    deg & km s$^{-1}$  &    kpc   \\

\hline
  1  &       I08076-3556  &  253.2952  &  -1.6159  &     5.9  $\pm$     0.4  &     1.0  $_{-     0.5  }  ^{+     0.5  }$  &   74  &       I16385-4619  &  338.5690  &  -0.1441  &  -117.8  $\pm$     0.4  &     7.0  $_{-     0.4  }  ^{+     0.3  }$  \\ 
  2  &       I08303-4303  &  261.6457  &  -2.0911  &    15.2  $\pm$     0.4  &     2.4  $_{-     0.6  }  ^{+     0.5  }$  &   75  &       I16424-4531  &  339.6226  &  -0.1225  &   -31.1  $\pm$     0.4  &     2.3  $_{-     0.2  }  ^{+     0.3  }$  \\ 
  3  &       I08448-4343  &  263.7731  &  -0.4337  &     3.4  $\pm$     0.4  &     1.4  $_{-     0.8  }  ^{+     0.6  }$  &   76  &       I16445-4459  &  340.2492  &  -0.0458  &  -121.4  $\pm$     0.4  &     8.2  $_{-     0.3  }  ^{+     0.3  }$  \\ 
  4  &       I08470-4243  &  263.2493  &   0.5114  &    13.5  $\pm$     0.4  &     2.4  $_{-     0.6  }  ^{+     0.6  }$  &   77  &       I16458-4512  &  340.2475  &  -0.3741  &   -49.9  $\pm$     0.4  &     3.5  $_{-     0.2  }  ^{+     0.2  }$  \\ 
  5  &       I09002-4732  &  268.4203  &  -0.8476  &     3.9  $\pm$     0.4  &     2.0  $_{-     0.8  }  ^{+     0.7  }$  &   78  &       I16484-4603  &  339.8842  &  -1.2572  &   -31.6  $\pm$     0.4  &     2.4  $_{-     0.2  }  ^{+     0.2  }$  \\ 
  6  &       I09018-4816  &  269.1529  &  -1.1308  &    10.8  $\pm$     0.4  &     2.8  $_{-     0.7  }  ^{+     0.6  }$  &   79  &       I16487-4423  &  341.2158  &  -0.2358  &   -43.2  $\pm$     0.4  &     3.3  $_{-     0.2  }  ^{+     0.2  }$  \\ 
  7  &       I09094-4803  &  269.8498  &  -0.0633  &    74.6  $\pm$     0.5  &     8.6  $_{-     0.4  }  ^{+     0.3  }$  &   80  &       I16489-4431  &  341.1258  &  -0.3475  &   -40.4  $\pm$     0.4  &     3.1  $_{-     0.2  }  ^{+     0.2  }$  \\ 
  8  &       I10365-5803  &  286.2076  &   0.1708  &   -18.9  $\pm$     0.4  &     3.4  $_{-     0.6  }  ^{+     0.6  }$  &   81  &       I16506-4512  &  340.7842  &  -1.0157  &   -25.8  $\pm$     0.4  &     2.1  $_{-     0.2  }  ^{+     0.2  }$  \\ 
  9  &       I11298-6155  &  293.8275  &  -0.7457  &    33.0  $\pm$     0.4  &     9.4  $_{-     0.3  }  ^{+     0.4  }$  &   82  &       I16524-4300  &  342.7076  &   0.1258  &   -40.6  $\pm$     0.4  &     3.2  $_{-     0.2  }  ^{+     0.2  }$  \\ 
 10  &       I11332-6258  &  294.5109  &  -1.6218  &   -16.1  $\pm$     0.5  &     1.1  $_{-     0.6  }  ^{+     0.7  }$  &   83  &       I16547-4247  &  343.1275  &  -0.0625  &   -29.7  $\pm$     0.4  &     2.6  $_{-     0.2  }  ^{+     0.2  }$  \\ 
 11  &       I11590-6452  &  297.7209  &  -2.7820  &    -3.9  $\pm$     0.4  &     0.4  $_{-     0.1  }  ^{+     0.5  }$  &   84  &       I16562-3959  &  345.4926  &   1.4686  &   -11.9  $\pm$     0.4  &     1.3  $_{-     0.2  }  ^{+     0.2  }$  \\ 
 12  &       I12320-6122  &  300.9692  &   1.1456  &   -42.5  $\pm$     0.4  &     3.1  $_{-     0.5  }  ^{+     0.6  }$  &   85  &       I16571-4029  &  345.1927  &   1.0288  &   -14.7  $\pm$     0.4  &     1.6  $_{-     0.2  }  ^{+     0.2  }$  \\ 
 13  &       I12326-6245  &  301.1358  &  -0.2258  &   -39.1  $\pm$     0.4  &     5.6  $_{-     0.8  }  ^{+     0.5  }$  &   86  &       I17006-4215  &  344.2208  &  -0.5941  &   -24.2  $\pm$     0.4  &     2.3  $_{-     0.2  }  ^{+     0.2  }$  \\ 
 14  &       I12383-6128  &  301.7308  &   1.1040  &   -37.9  $\pm$     0.4  &     2.8  $_{-     0.6  }  ^{+     0.7  }$  &   87  &       I17008-4040  &  345.5043  &   0.3475  &   -15.9  $\pm$     0.4  &     1.7  $_{-     0.2  }  ^{+     0.2  }$  \\ 
 15  &       I12572-6316  &  303.9308  &  -0.6874  &    29.5  $\pm$     0.4  &    11.5  $_{-     0.3  }  ^{+     0.3  }$  &   88  &       I17016-4124  &  345.0029  &  -0.2241  &   -26.5  $\pm$     0.4  &     2.5  $_{-     0.2  }  ^{+     0.2  }$  \\ 
 16  &       I13079-6218  &  305.2091  &   0.2058  &   -40.5  $\pm$     0.4  &     3.0  $_{-     0.7  }  ^{+     0.6  }$  &   89  &       I17136-3617  &  350.5074  &   0.9590  &   -10.3  $\pm$     0.4  &     1.6  $_{-     0.2  }  ^{+     0.3  }$  \\ 
 17  &       I13080-6229  &  305.1958  &   0.0342  &   -34.0  $\pm$     0.4  &     2.3  $_{-     0.6  }  ^{+     0.6  }$  &   90  &       I17143-3700  &  350.0158  &   0.4325  &   -31.3  $\pm$     0.4  &    12.3  $_{-     0.2  }  ^{+     0.2  }$  \\ 
 18  &       I13111-6228  &  305.5624  &   0.0142  &   -38.7  $\pm$     0.4  &     2.7  $_{-     0.6  }  ^{+     0.7  }$  &   91  &       I17158-3901  &  348.5492  &  -0.9789  &   -15.8  $\pm$     0.4  &     2.0  $_{-     0.2  }  ^{+     0.2  }$  \\ 
 19  &       I13134-6242  &  305.7989  &  -0.2441  &   -31.6  $\pm$     0.4  &     2.0  $_{-     0.4  }  ^{+     0.6  }$  &   92  &       I17160-3707  &  350.1025  &   0.0825  &   -69.7  $\pm$     0.4  &     9.9  $_{-     0.2  }  ^{+     0.2  }$  \\ 
 20  &       I13140-6226  &  305.8872  &   0.0158  &   -32.7  $\pm$     0.4  &     2.1  $_{-     0.5  }  ^{+     0.6  }$  &   93  &       I17175-3544  &  351.4159  &   0.6457  &    -5.7  $\pm$     0.4  &     1.0  $_{-     0.3  }  ^{+     0.3  }$  \\ 
 21  &       I13291-6229  &  307.6158  &  -0.2592  &   -37.1  $\pm$     0.4  &     2.4  $_{-     0.5  }  ^{+     0.6  }$  &   94  &       I17204-3636  &  351.0411  &  -0.3357  &   -17.8  $\pm$     0.4  &     2.7  $_{-     0.2  }  ^{+     0.2  }$  \\ 
 22  &       I13291-6249  &  307.5608  &  -0.5875  &   -32.4  $\pm$     0.4  &     7.8  $_{-     0.5  }  ^{+     0.5  }$  &   95  &       I17220-3609  &  351.5809  &  -0.3524  &   -95.0  $\pm$     0.4  &     7.7  $_{-     0.2  }  ^{+     0.2  }$  \\ 
 23  &       I13295-6152  &  307.7558  &   0.3525  &   -43.5  $\pm$     0.5  &     2.9  $_{-     0.6  }  ^{+     0.7  }$  &   96  &       I17233-3606  &  351.7742  &  -0.5374  &    -4.2  $\pm$     0.4  &     0.7  $_{-     0.3  }  ^{+     0.3  }$  \\ 
 24  &       I13471-6120  &  309.9208  &   0.4775  &   -57.3  $\pm$     0.4  &     6.6  $_{-     0.6  }  ^{+     0.6  }$  &   97  &       I17244-3536  &  352.3158  &  -0.4425  &    -9.4  $\pm$     0.4  &     1.8  $_{-     0.3  }  ^{+     0.3  }$  \\ 
 25  &       I13484-6100  &  310.1442  &   0.7591  &   -54.1  $\pm$     0.4  &     6.9  $_{-     0.7  }  ^{+     0.5  }$  &   98  &       I17258-3637  &  351.6326  &  -1.2538  &   -11.7  $\pm$     0.4  &     2.0  $_{-     0.3  }  ^{+     0.2  }$  \\ 
 26  &       I14013-6105  &  311.6257  &   0.2891  &   -54.6  $\pm$     0.4  &     3.5  $_{-     0.5  }  ^{+     0.7  }$  &   99  &       I17269-3312  &  354.5959  &   0.4691  &   -20.7  $\pm$     0.5  &     4.2  $_{-     0.2  }  ^{+     0.2  }$  \\ 
 27  &       I14050-6056  &  312.1078  &   0.3091  &   -48.1  $\pm$     0.4  &     3.1  $_{-     0.5  }  ^{+     0.5  }$  &  100  &       I17271-3439  &  353.4091  &  -0.3607  &   -15.9  $\pm$     0.6  &     3.0  $_{-     0.2  }  ^{+     0.2  }$  \\ 
 28  &       I14164-6028  &  313.5758  &   0.3242  &   -45.2  $\pm$     0.4  &     2.8  $_{-     0.4  }  ^{+     0.4  }$  &  101  &       I17278-3541  &  352.6308  &  -1.0674  &     0.7  $\pm$     0.4  &     1.3  $_{-     0.0  }  ^{+     0.2  }$  \\ 
 29  &       I14206-6151  &  313.5758  &  -1.1540  &   -48.8  $\pm$     0.4  &     3.0  $_{-     0.5  }  ^{+     0.5  }$  &  102  &       I17439-2845  &    0.3160  &  -0.2008  &    18.8  $\pm$     0.4  &     8.1  $_{-     0.0  }  ^{+     0.0  }$  \\ 
 30  &       I14212-6131  &  313.7658  &  -0.8624  &   -49.2  $\pm$     0.4  &     3.1  $_{-     0.5  }  ^{+     0.5  }$  &  103  &       I17441-2822  &    0.6659  &  -0.0341  &    51.5  $\pm$     1.6  &     8.1  $_{-     0.6  }  ^{+     0.6  }$  \\ 
 31  &       I14382-6017  &  316.1392  &  -0.5058  &   -60.3  $\pm$     0.4  &     8.1  $_{-     0.6  }  ^{+     0.4  }$  &  104  &       I17455-2800  &    1.1258  &  -0.1075  &   -14.2  $\pm$     0.4  &    10.0  $_{-     0.7  }  ^{+     0.7  }$  \\ 
 32  &       I14453-5912  &  317.4074  &   0.1092  &   -40.2  $\pm$     0.4  &     2.3  $_{-     0.3  }  ^{+     0.3  }$  &  105  &       I17545-2357  &    5.6374  &   0.2375  &     9.2  $\pm$     0.4  &     2.4  $_{-     0.3  }  ^{+     0.2  }$  \\ 
 33  &       I14498-5856  &  318.0495  &   0.0858  &   -49.7  $\pm$     0.4  &     2.9  $_{-     0.3  }  ^{+     0.4  }$  &  106  &       I17589-2312  &    6.7959  &  -0.2575  &    21.6  $\pm$     0.5  &     3.7  $_{-     0.2  }  ^{+     0.2  }$  \\ 
 34  &       I15122-5801  &  321.0545  &  -0.5073  &   -60.3  $\pm$     0.4  &     9.2  $_{-     0.4  }  ^{+     0.3  }$  &  107  &       I17599-2148  &    8.1408  &   0.2225  &    19.8  $\pm$     0.4  &     3.1  $_{-     0.2  }  ^{+     0.2  }$  \\ 
 35  &       I15254-5621  &  323.4591  &  -0.0791  &   -66.9  $\pm$     0.4  &     3.8  $_{-     0.3  }  ^{+     0.4  }$  &  108  &       I18032-2032  &    9.6209  &   0.1942  &     5.9  $\pm$     0.4  &     5.2  $_{-     0.3  }  ^{+     0.2  }$  \\ 
 36  &       I15290-5546  &  324.2011  &   0.1208  &   -88.5  $\pm$     0.4  &     8.1  $_{-     0.6  }  ^{+     0.4  }$  &  109  &       I18056-1952  &   10.4725  &   0.0275  &    66.7  $\pm$     0.4  &    10.2  $_{-     0.2  }  ^{+     0.1  }$  \\ 
 37  &       I15384-5348  &  326.4457  &   0.9073  &   -40.6  $\pm$     0.4  &     2.4  $_{-     0.3  }  ^{+     0.3  }$  &  110  &       I18075-2040  &   10.0008  &  -0.7541  &    31.5  $\pm$     0.4  &     3.8  $_{-     0.2  }  ^{+     0.2  }$  \\ 
 38  &       I15394-5358  &  326.4741  &   0.7024  &   -40.7  $\pm$     0.4  &     2.4  $_{-     0.2  }  ^{+     0.3  }$  &  111  &       I18079-1756  &   12.4176  &   0.5058  &    18.4  $\pm$     0.4  &     2.2  $_{-     0.2  }  ^{+     0.2  }$  \\ 
 39  &       I15408-5356  &  326.6573  &   0.5940  &   -40.1  $\pm$     0.4  &     2.4  $_{-     0.3  }  ^{+     0.3  }$  &  112  &       I18089-1732  &   12.8875  &   0.4891  &    33.5  $\pm$     0.4  &     3.4  $_{-     0.2  }  ^{+     0.2  }$  \\ 
 40  &       I15411-5352  &  326.7240  &   0.6140  &   -41.0  $\pm$     0.4  &     2.4  $_{-     0.3  }  ^{+     0.2  }$  &  113  &       I18110-1854  &   11.9355  &  -0.6156  &    39.0  $\pm$     0.4  &     3.9  $_{-     0.2  }  ^{+     0.2  }$  \\ 
 41  &       I15437-5343  &  327.1194  &   0.5090  &   -83.8  $\pm$     0.5  &     4.7  $_{-     0.3  }  ^{+     0.4  }$  &  114  &       I18116-1646  &   13.8725  &   0.2808  &    50.2  $\pm$     0.4  &     4.3  $_{-     0.1  }  ^{+     0.2  }$  \\ 
 42  &       I15439-5449  &  326.4724  &  -0.3774  &   -53.4  $\pm$     0.4  &     3.1  $_{-     0.3  }  ^{+     0.3  }$  &  115  &       I18117-1753  &   12.9075  &  -0.2591  &    38.0  $\pm$     0.4  &     3.7  $_{-     0.2  }  ^{+     0.2  }$  \\ 
 43  &       I15502-5302  &  328.3075  &   0.4309  &   -91.6  $\pm$     0.4  &     5.3  $_{-     0.4  }  ^{+     0.4  }$  &  116  &       I18134-1942  &   11.4974  &  -1.4853  &    10.7  $\pm$     0.4  &     1.5  $_{-     0.2  }  ^{+     0.2  }$  \\ 
 44  &       I15520-5234  &  328.8092  &   0.6325  &   -41.2  $\pm$     0.4  &     2.5  $_{-     0.2  }  ^{+     0.3  }$  &  117  &       I18139-1842  &   12.4310  &  -1.1138  &    40.0  $\pm$     0.4  &     3.9  $_{-     0.2  }  ^{+     0.2  }$  \\ 
 45  &       I15522-5411  &  327.8092  &  -0.6358  &   -46.8  $\pm$     0.4  &     2.8  $_{-     0.3  }  ^{+     0.2  }$  &  118  &       I18159-1648  &   14.3308  &  -0.6440  &    23.9  $\pm$     0.4  &     2.5  $_{-     0.2  }  ^{+     0.2  }$  \\ 
 46  &       I15557-5215  &  329.4691  &   0.5024  &   -67.6  $\pm$     0.4  &     3.9  $_{-     0.3  }  ^{+     0.3  }$  &  119  &       I18182-1433  &   16.5858  &  -0.0508  &    60.1  $\pm$     0.4  &     4.4  $_{-     0.2  }  ^{+     0.2  }$  \\ 
 47  &       I15567-5236  &  329.3374  &   0.1475  &  -107.3  $\pm$     0.4  &     6.2  $_{-     0.4  }  ^{+     0.4  }$  &  120  &       I18223-1243  &   18.6542  &  -0.0591  &    46.0  $\pm$     0.4  &     3.4  $_{-     0.2  }  ^{+     0.2  }$  \\ 
 48  &       I15570-5227  &  329.4724  &   0.2158  &  -100.2  $\pm$     0.4  &     5.7  $_{-     0.4  }  ^{+     0.5  }$  &  121  &       I18228-1312  &   18.3010  &  -0.3891  &    33.6  $\pm$     0.4  &     2.7  $_{-     0.2  }  ^{+     0.2  }$  \\ 
 49  &       I15584-5247  &  329.4241  &  -0.1624  &   -76.2  $\pm$     0.4  &     4.3  $_{-     0.3  }  ^{+     0.3  }$  &  122  &       I18236-1205  &   19.3625  &  -0.0308  &    27.8  $\pm$     0.4  &     2.2  $_{-     0.2  }  ^{+     0.2  }$  \\ 
 50  &       I15596-5301  &  329.4058  &  -0.4591  &   -73.9  $\pm$     0.4  &     9.9  $_{-     0.3  }  ^{+     0.3  }$  &  123  &       I18264-1152  &   19.8825  &  -0.5341  &    45.4  $\pm$     0.4  &     3.3  $_{-     0.2  }  ^{+     0.2  }$  \\ 

\hline
\end{tabular}
}
\end{table*}

\begin{table*}
	\centering
	\contcaption{Distances of 146 ATOMS clumps.}
	\label{atomsample}
	\resizebox{18cm}{!}{
	\begin{tabular}{cccccccccccc} 
		\hline

 ID &             Name &              $l$  &             $b$ &   $V_{\rm lsr}$  &             $D$  &  ID &             Name &              $l$  &             $b$ &   $V_{\rm lsr}$  &             $D$  \\
    &                  &                 deg  &    deg & km s$^{-1}$  &    kpc   &     &                  &                 deg  &    deg & km s$^{-1}$  &    kpc   \\

\hline
 51  &       I16026-5035  &  331.3608  &   1.0657  &   -78.5  $\pm$     0.4  &     4.5  $_{-     0.3  }  ^{+     0.2  }$  &  124  &       I18290-0924  &  22.3508  &   0.0675  &    84.3  $\pm$     0.4  &     5.0  $_{-     0.2  }  ^{+     0.2  }$  \\ 
 52  &       I16037-5223  &  330.2943  &  -0.3941  &   -81.0  $\pm$     0.4  &     9.6  $_{-     0.3  }  ^{+     0.3  }$  &  125  &       I18308-0503  &  26.4212  &   1.6902  &    43.6  $\pm$     0.4  &     2.8  $_{-     0.2  }  ^{+     0.2  }$  \\ 
 53  &       I16060-5146  &  330.9542  &  -0.1825  &   -88.8  $\pm$     0.4  &     5.0  $_{-     0.3  }  ^{+     0.4  }$  &  126  &       I18311-0809  &  23.7107  &   0.1708  &   113.5  $\pm$     0.4  &     6.6  $_{-     0.3  }  ^{+     0.4  }$  \\ 
 54  &       I16065-5158  &  330.8792  &  -0.3675  &   -61.5  $\pm$     0.4  &     3.6  $_{-     0.3  }  ^{+     0.2  }$  &  127  &       I18314-0720  &  24.4709  &   0.4874  &   102.1  $\pm$     0.5  &     5.9  $_{-     0.3  }  ^{+     0.4  }$  \\ 
 55  &       I16071-5142  &  331.1325  &  -0.2442  &   -85.9  $\pm$     0.5  &     4.9  $_{-     0.3  }  ^{+     0.3  }$  &  128  &       I18316-0602  &  25.6492  &   1.0507  &    45.3  $\pm$     0.4  &     2.9  $_{-     0.2  }  ^{+     0.2  }$  \\ 
 56  &       I16076-5134  &  331.2792  &  -0.1891  &   -86.0  $\pm$     0.4  &     4.9  $_{-     0.3  }  ^{+     0.3  }$  &  129  &       I18317-0513  &  26.3808  &   1.4071  &    42.2  $\pm$     0.4  &     2.7  $_{-     0.2  }  ^{+     0.2  }$  \\ 
 57  &       I16119-5048  &  332.2958  &  -0.0941  &   -48.4  $\pm$     0.4  &     3.0  $_{-     0.3  }  ^{+     0.2  }$  &  130  &       I18317-0757  &  23.9539  &   0.1508  &    80.5  $\pm$     0.4  &     4.7  $_{-     0.2  }  ^{+     0.3  }$  \\ 
 58  &       I16132-5039  &  332.5440  &  -0.1241  &   -47.0  $\pm$     0.4  &     2.9  $_{-     0.2  }  ^{+     0.3  }$  &  131  &       I18341-0727  &  24.6725  &  -0.1508  &   113.5  $\pm$     0.4  &     6.6  $_{-     0.3  }  ^{+     0.4  }$  \\ 
 59  &       I16158-5055  &  332.6474  &  -0.6090  &   -49.4  $\pm$     0.4  &     3.0  $_{-     0.2  }  ^{+     0.2  }$  &  132  &       I18411-0338  &  28.8608  &   0.0659  &   103.7  $\pm$     0.4  &     8.3  $_{-     0.5  }  ^{+     0.3  }$  \\ 
 60  &       I16164-5046  &  332.8256  &  -0.5490  &   -56.2  $\pm$     0.4  &     3.4  $_{-     0.2  }  ^{+     0.2  }$  &  133  &       I18434-0242  &  29.9538  &  -0.0158  &    98.1  $\pm$     0.4  &     5.7  $_{-     0.4  }  ^{+     0.4  }$  \\ 
 61  &       I16172-5028  &  333.1344  &  -0.4307  &   -52.1  $\pm$     0.4  &     3.2  $_{-     0.2  }  ^{+     0.2  }$  &  134  &       I18440-0148  &  30.8175  &   0.2742  &    98.3  $\pm$     0.4  &     5.7  $_{-     0.4  }  ^{+     0.5  }$  \\ 
 62  &       I16177-5018  &  333.3077  &  -0.3657  &   -50.9  $\pm$     0.4  &     3.1  $_{-     0.2  }  ^{+     0.2  }$  &  135  &       I18445-0222  &  30.3860  &  -0.1041  &    87.2  $\pm$     0.4  &     5.0  $_{-     0.3  }  ^{+     0.3  }$  \\ 
 63  &       I16272-4837  &  335.5857  &  -0.2908  &   -45.8  $\pm$     0.4  &     3.0  $_{-     0.2  }  ^{+     0.2  }$  &  136  &       I18461-0113  &  31.5808  &   0.0775  &    97.1  $\pm$     0.4  &     5.6  $_{-     0.4  }  ^{+     0.5  }$  \\ 
 64  &       I16297-4757  &  336.3593  &  -0.1374  &   -79.9  $\pm$     0.4  &     4.7  $_{-     0.2  }  ^{+     0.2  }$  &  137  &       I18469-0132  &  31.3958  &  -0.2575  &    86.9  $\pm$     0.4  &     5.2  $_{-     0.3  }  ^{+     0.4  }$  \\ 
 65  &       I16304-4710  &  337.0042  &   0.3225  &   -61.0  $\pm$     0.4  &    11.2  $_{-     0.2  }  ^{+     0.2  }$  &  138  &       I18479-0005  &  32.7973  &   0.1908  &    15.1  $\pm$     0.4  &    12.8  $_{-     0.2  }  ^{+     0.3  }$  \\ 
 66  &       I16313-4729  &  336.8658  &   0.0042  &   -72.3  $\pm$     0.4  &     4.4  $_{-     0.2  }  ^{+     0.2  }$  &  139  &       I18502+0051  &  34.2601  &  -0.4259  &    53.0  $\pm$     0.4  &    10.5  $_{-     0.3  }  ^{+     0.3  }$  \\ 
 67  &       I16318-4724  &  336.9942  &  -0.0275  &  -120.5  $\pm$     0.4  &     8.1  $_{-     0.3  }  ^{+     0.3  }$  &  140  &       I18507+0110  &  34.2575  &   0.1542  &    58.5  $\pm$     0.4  &     1.6  $_{-     0.3  }  ^{+     0.3  }$  \\ 
 68  &       I16330-4725  &  337.1208  &  -0.1741  &   -74.1  $\pm$     0.4  &    10.5  $_{-     0.2  }  ^{+     0.2  }$  &  141  &       I18507+0121  &  34.4108  &   0.2342  &    58.2  $\pm$     0.4  &     3.4  $_{-     0.3  }  ^{+     0.3  }$  \\ 
 69  &       I16344-4658  &  337.6125  &  -0.0592  &   -48.9  $\pm$     0.4  &    11.8  $_{-     0.2  }  ^{+     0.2  }$  &  142  &       I18517+0437  &  37.4308  &   1.5188  &    44.1  $\pm$     0.4  &     2.5  $_{-     0.2  }  ^{+     0.4  }$  \\ 
 70  &       I16348-4654  &  337.7042  &  -0.0541  &   -47.2  $\pm$     0.4  &    11.9  $_{-     0.2  }  ^{+     0.2  }$  &  143  &       I18530+0215  &  35.4657  &   0.1408  &    77.8  $\pm$     0.4  &     4.5  $_{-     0.4  }  ^{+     0.4  }$  \\ 
 71  &       I16351-4722  &  337.4058  &  -0.4025  &   -40.3  $\pm$     0.4  &     2.8  $_{-     0.2  }  ^{+     0.2  }$  &  144  &       I19078+0901  &  43.1658  &   0.0109  &     7.2  $\pm$     0.4  &    11.5  $_{-     0.3  }  ^{+     0.3  }$  \\ 
 72  &       I16362-4639  &  338.0642  &  -0.0675  &   -37.4  $\pm$     0.4  &     2.7  $_{-     0.2  }  ^{+     0.2  }$  &  145  &       I19095+0930  &  43.7941  &  -0.1275  &    44.7  $\pm$     0.4  &     9.2  $_{-     0.4  }  ^{+     0.3  }$  \\ 
 73  &       I16372-4545  &  338.8506  &   0.4091  &   -57.5  $\pm$     0.4  &     3.8  $_{-     0.2  }  ^{+     0.2  }$  &  146  &       I19097+0847  &  43.1791  &  -0.5191  &    57.3  $\pm$     0.4  &     8.5  $_{-     0.4  }  ^{+     0.5  }$  \\ 

\hline
\end{tabular}
}

\begin{flushleft}
{ NOTE:}  I17441-2822 and 17455-2800 are close to the Galactic centre, whose distance values are adopted from \citet{Fau04}; I18469-0132, I18507+0110, and I19097+0847 have the maser parallax distance measurements \citep{Rei14}.

\end{flushleft}
\end{table*}

\section{Calculation of mass and mass surface density}
\label{sec:app-mass} 
To understand the nature of the 90 COM-containing cores not associated the \chii\ signature (see Sect.\,\ref{sec:core:classify-HMC}) and 274 ``unknown" cores, 
we calculated their mass $M_{core}$ in the following form:
\begin{equation}
\label{eqn:mass}
M_{\rm{core}} \, = \, \frac{F_\nu^{\rm int} \, R_{\rm gd} \, D^2 }{B_\nu(T_{\rm{dust}}) \, \kappa_\nu},
\end{equation}
where $F_\nu^{\rm int}$ is the measured integrated 3\,mm flux over the core size,  $R_{\rm gd}$ is the gas-to-dust mass ratio (assumed to be 100), $D$ is the distance to the core,
$B_\nu$ is the Planck function for a dust temperature $T_{\rm{dust}}$, and $\kappa_\nu$ is the opacity assumed to be $0.18$\,cm$^2$\,g$^{-1}$ of dust at $\nu \sim94$\,GHz, taken directly from the ``OH5" dust model, a combination of dust from \citet{Oss94} and \citet{Pol94} and extended to longer wavelengths by \citet{You05}.
Moreover, the mass surface density $\Sigma_{core}$ was derived from $\Sigma_{\rm core} = M_{\rm core}/(\pi R_{\rm core}^2)$,
where $R_{\rm core}$ is the core radius equal to the geometric mean of the $FWHM_{\rm maj}^{\rm dec}$, $FWHM_{\rm min}^{\rm dec}$ at the respective core distance.
Given the median radius of the 90 COM-containing cores $\sim 0.02$\,pc, the $T_{\rm{dust}}\sim 100$\,K was assumed for a first-order approximation
in the calculation, which is typical of HMCs at a radius of $\sim 0.01$\,pc \citep{Oso99}. 
For the 274 ``unknown" cores, the temperature 25\,K is assumed, which is comparable to the mean value over the clumps that do not have cHMC or \chii\ signatures.

The peak column densities are estimated from the peak flux density of the cores using the following equation:
\begin{equation}
N_{\rm{H_2}} \, = \, \frac{F_\nu^{\rm p} \, R_{\rm gd}}{B_\nu(T_{\rm{dust}}) \, \Omega \, \kappa_\nu \, \umu\, m_{\rm{H}}},
\end{equation}
where $\Omega$ is the beam solid angle, $\umu$ is the mean molecular weight of the gas, assumed to be 2.8, $m_{\rm{H}}$ is the mass of the hydrogen atom. In turn, the number 
densities can be estimated via $n_{\rm{H_2}}=N_{\rm{H_2}}/2R$, assuming a spherical geometry for cores.

The three derived parameters $M_{core}$, $n_{\rm{H_2}}$, and $\Sigma_{core}$ are arranged in last three columns of Tables\,\ref{tbl:hmc-candidate} for the 32 \CHMC s, and of Table\,\ref{tbl:hmc-potential} for 58 \PCHMC s.


\vspace{5mm}
\noindent
Author affiliations:\\

\noindent 
$^{1}$Department of Astronomy, Yunnan University, Kunming, 650091, PR China \\
$^{2}$Shanghai Astronomical Observatory, Chinese Academy of Sciences, 80 Nandan Road, Shanghai 200030, Peoples Republic of China \\
$^{3}$Key Laboratory for Research in Galaxies and Cosmology, Shanghai Astronomical Observatory, Chinese Academy of Sciences, 80 Nandan Road, Shanghai 200030, Peoples Republic of China \\
$^{4}$Departamento de Astronom\'ia, Universidad de Concepci\'on, Av. Esteban Iturra s/n, Distrito Universitario, 160-C, Chile \\
$^{5}$Department of Astronomy, The University of Texas at Austin, 2515 Speedway, Stop C1400, Austin, TX 78712-1205, USA\\
$^{6}$Korea Astronomy and Space Science Institute, 776 Daedeokdaero, Yuseong-gu, Daejeon 34055, Republic of Korea\\
$^{7}$Kavli Institute for Astronomy and Astrophysics, Peking University, 5 Yiheyuan Road, Haidian District, Beijing 100871, People's Republic of China\\
$^{8}$Departamento de Astronom\'{\i}a, Universidad de Chile, Las Condes, Santiago, Chile\\
$^{9}$Max-Planck-Institute for Astronomy, K\"{o}nigstuhl 17, 69117 Heidelberg, Germany \\
$^{10}$Jet Propulsion Laboratory, California Institute of Technology, 4800 Oak Grove Drive, Pasadena, CA 91109, USA\\
$^{11}$Institute of Astronomy and Astrophysics, Academia Sinica. 11F of Astronomy-Mathematics Building, AS/NTU No. 1, Section 4, Roosevelt Rd., Taipei 10617, Taiwan\\
$^{12}$Indian Institute of Space Science and Technology, Thiruvananthapuram 695 547, Kerala, India\\
$^{13}$Center for Astrophysics $|$ Harvard \& Smithsonian, 60 Garden Street, Cambridge, MA 02138, USA\\
$^{14}$Department of Physics, P.O. Box 64, FI-00014, University of Helsinki, Finland\\
$^{15}$National Astronomical Observatories, Chinese Academy of Sciences, Beijing 100101, China  \\
$^{16}$University of Chinese Academy of Sciences, Beijing 100049, China\\
$^{17}$NAOC-UKZN Computational Astrophysics Centre, University of KwaZulu-Natal, Durban 4000, South Africa\\
$^{18}$Key Laboratory of Radio Astronomy, Chinese Academy of Sciences, Nanjing 210008, People's Republic of China\\
$^{19}$Department of Astronomy, Peking University, 100871, Beijing, People's Republic of China\\
$^{20}$University of Science and Technology, Korea (UST), 217 Gajeong-ro, Yuseong-gu, Daejeon 34113, Republic of Korea\\
$^{21}$National Astronomical Observatory of Japan, National Institutes of Natural Sciences, 2-21-1 Osawa, Mitaka, Tokyo 181-8588, Japan\\
$^{22}$School of Physics, University of New South Wales, Sydney, NSW 2052, Australia\\
$^{23}$School of Space Research, Kyung Hee University, Yongin-Si, Gyeonggi-Do 17104, Republic of Korea\\
$^{24}$Astronomy Department, University of California, Berkeley, CA 94720, USA\\
$^{25}$IRAP, Universit\'{e} de Toulouse, CNRS, UPS, CNES, Toulouse, France\\
$^{26}$School of Physics and Astronomy, Sun Yat-sen University, 2 Daxue Road, Zhuhai, Guangdong, 519082, People's Republic of China\\
$^{27}$E\"{o}tv\"{o}s Lor\'{a}nd University, Department of Astronomy, P\'{a}zm\'{a}ny P\'{e}ter s\'{e}t\'{a}ny 1/A, H-1117, Budapest, Hungary\\
$^{28}$College of Science, Yunnan Agricultural University, Kunming, 650201, People's Republic of China\\
$^{29}$Satyendra Nath Bose National Centre for Basic Sciences, Block-JD, Sector-III, Salt Lake, Kolkata-700 106 \\
$^{30}$Physical Research Laboratory, Navrangpura, Ahmedabad—380 009, India \\
$^{31}$Department of Physics, Taiyuan Normal University, Jinzhong 030619, China \\
$^{32}$SOFIA Science Centre, USRA, NASA Ames Research Centre, MS-12, N232, Moffett Field, CA 94035, USA \\

\label{lastpage}
\end{document}